\documentclass[11pt]{article}
\AtBeginDocument{%
  \providecommand\BibTeX{{%
    \normalfont B\kern-0.5em{\scshape i\kern-0.25em b}\kern-0.8em\TeX}}}

\usepackage{tabularx}
\usepackage{graphicx}
\usepackage{adjustbox}
\usepackage{algorithm}
\usepackage{algpseudocode}
\usepackage{amsmath}
\usepackage{multirow}
\usepackage{makecell}
\usepackage[super]{nth}

\usepackage{todonotes}
\usepackage{pgfplots}
\usepackage{pgfplotstable}
\usepackage{tikz}
\pgfplotsset{compat=1.7}
\usepackage[numbers]{natbib}
\usepackage{soul}
\usepackage{subcaption}
\usepackage{caption}
\usepackage{tikz}
\usepackage{collcell}
\usepackage{booktabs}
\usepackage{siunitx}
\usepackage[toc,page]{appendix}
\usepackage{float}
\usepackage{comment}

\usepgfplotslibrary{groupplots}
\pgfplotsset{compat=1.8}
\usepgfplotslibrary{statistics}

\newcolumntype{M}[1]{>{\centering\arraybackslash}m{#1}}
\newcolumntype{L}[1]{>{\raggedright\arraybackslash}p{#1}}

\makeatletter
\newcommand\avsuminner[2]{%
  {\sbox0{$\m@th#1\sum$}%
   \vphantom{\usebox0}%
   \ooalign{%
     \hidewidth
     \smash{\vrule height\dimexpr\ht0+1pt\relax depth\dimexpr\dp0+1pt\relax}%
     \hidewidth\cr
     $\m@th#1\sum$\cr
   }%
  }%
}
\makeatother

\newcolumntype{P}[1]{>{\centering\arraybackslash}p{#1}}
\newcommand{\testcol}{\textit{GenTREC}}

\title{\testcol: The First  Test Collection Generated by Large Language Models for Evaluating Information Retrieval Systems} % tech report

 \author{Mehmet Deniz Türkmen$^1$, Mucahid Kutlu$^2$, \\ Bahadir Altun$^3$, Gokalp Cosgun$^4$ \\
  $^1$GESIS Leibniz-Institut für Sozialwissenschaften, 
  $^2$Qatar University \\  
  $^3$University at Buffalo, 
  $^4$Wayne State University \\  
  \texttt{deniz.tuerkmen@gesis.org},
  \texttt{mucahidkutlu@qu.edu.qa}, \\ \texttt{ialtun@buffalo.edu}, 
  \texttt{gcosgun@wayne.edu}
\\}

\begin{document}

%%
%% The "title" command has an optional parameterC

%\author{Mehmet Deniz Türkmen$^{a}$\thanks{corresponding author}}
%\author{Mucahid Kutlu$^{b}$}
%\author{Bahadir Altun$^{c}$}
%\author{\\Gokalp Cosgun$^{d}$}

%\affiliation{organization={GESIS Leibniz-Institut für Sozialwissenschaften, KTS Department}, city={Cologne}, country={Germany}}

%\affiliation{organization={Qatar University, Department of Computer Science and Engineering}, city={Doha}, country={Qatar}}

%\affiliation{organization={University at Buffalo, Computer Science and Engineering Department}, city={Buffalo}, country={USA}}

%\affiliation{organization={Wayne State University,Industrial and Systems Engineering}, city={Detroit}, country={USA}}

%\affiliation{
%  \institution{Dept.\ of Computer Engineering, TOBB U.\ of Economics Technology}
%  \country{Turkey}
%}

\maketitle

\begin{abstract}

Building test collections for Information Retrieval evaluation has traditionally been a resource-intensive and time-consuming task, primarily due to the dependence on manual relevance judgments. While various cost-effective strategies have been explored, the development of such collections remains a significant challenge. In this paper, we present \testcol{}, the first test collection constructed entirely from documents generated by a Large Language Model (LLM), eliminating the need for manual relevance judgments. Our approach is based on the assumption that documents generated by an LLM are inherently relevant to the prompts used for their generation. Based on this heuristic, we utilized existing TREC search topics to generate documents. We consider a document relevant only to the prompt that generated it, while other document-topic pairs are treated as non-relevant. To introduce realistic retrieval challenges, we also generated non-relevant documents, ensuring that IR systems are tested against a diverse and robust set of materials. The resulting \testcol{} collection comprises 96,196 documents, 300 topics, and 18,964 relevance ``judgments". We conducted extensive experiments to evaluate \testcol{} in terms of document quality, relevance judgment accuracy, and evaluation reliability. Notably, our findings indicate that the ranking of IR systems using \testcol{} is compatible with the evaluations conducted using traditional TREC test collections, particularly for  P@100, MAP, and RPrec metrics. Overall, our results show that our proposed approach offers a promising, low-cost alternative for IR evaluation,  significantly reducing the burden of building and maintaining future IR evaluation resources.

%The results demonstrate that GenTC can serve as a promising, cost-effective alternative to traditional test collections. While it requires no manual annotation and offers a highly scalable solution, it still exhibits strong comparability to established human-generated resources. Our findings indicate that GenTC not only reduces the labor and expense of test collection creation but may also broaden the horizons for IR evaluation, enabling more frequent, flexible, and adaptive benchmarking as the field continues to evolve.

\end{abstract}
% part of the formatted document.
%\maketitle

\section{Introduction}\label{sec:intro}

%The construction of test collections for evaluating information retrieval (IR) systems has been a fundamental component of IR research, enabling the systematic comparison of various retrieval techniques and algorithms. Over the years, researchers have explored several to develop high-quality and reusable test collection with a reasonable budget. While reducing its cost is a highly challenging task, the conventional approach to developing a test collection is also risky because its quality can only be assessed once the collection is fully constructed. Specifically, the conventional way to develop a test collection consists of crawling a large  document collection, creating a set of search topics, selecting documents to be judged and collecting relevance judgments for the corresponding document-topic pairs. However, during the document collection process we have limited control over the resultant data. Thus, developing a search topic becomes a challenging task because there might be no or too many  relevant document for a particular topic. Moreover, in the standard pooling technique, the documents to be judged eventually rely on the systems that participated in the pool. Therefore, despite the well-considered design decisions implemented during their development, there are test collections in the literature that are of low quality and lack reusability [REF]. This results in  the substantial costs and valuable human efforts invested in their creation ultimately futile. 
The construction of test collections to evaluate IR systems has been a fundamental component of IR research, enabling systematic comparison of various retrieval techniques and algorithms. Typically, the process involves crawling a large document collection, developing a set of search topics, selecting documents for relevance judgment, and collecting relevance judgments for the document-topic pairs \cite{sanderson2010test}.  Over the years, researchers have developed several methodologies aimed at constructing high-quality and reusable test collections within reasonable budgets \cite{rahman2020efficient,sakai2007alternatives,jones1975report, cormack2018beyond,kutlu2018intelligent}. 

Several factors affect the quality of a test collection, including the number of topics, the accuracy and prevalence of relevance judgments, and the documents themselves. In particular, several studies explore the impact of the number of topics on the evaluation reliability and report that 50 topics, as used by several TREC test collections \cite{trec8-overview,harman1996overview}, are not enough for a reliable evaluation \cite{sakai2016topic,urbano2013measurement,voorhees2009topic,jones1975report}. In addition, the document collection should contain a reasonable amount of relevant documents for each topic to assess and distinguish the performance of systems. Furthermore, as collecting relevance judgments is extremely costly, a sample of the documents should be selected and judged. Obviously, the accuracy of the judgments and the selected documents to be judged \cite{altun2020building} affect the evaluation reliability. Considering all these factors affecting the quality of test collections,  the conventional approach to test collection construction carries inherent risks because the quality of a collection can only be fully assessed after its completion. As a result, despite careful design, test collections might suffer from low quality  \cite{voorhees2018building}, leading to substantial financial costs and considerable human effort being wasted.

In recent years, the remarkable success of large language models (LLMs) has also affected the IR research community in several directions.  In particular, a growing portion of document collections on the Internet is likely to consist of AI-generated content\footnote{https://www.technologyreview.com/2022/12/20/1065667/how-ai-generated-text-is-poisoning-the-internet/}. Therefore, IR systems needs to deal with this new AI-generated data. In addition, several researchers explored various ways to integrate these models into search operations \cite{zhai2024large}. While LLMs can be used to retrieve answers to a particular question, their knowledge base is restricted by the training data, limiting their usage to get information for recent events. Consequently, prior work investigated hybrid systems where a response is generated from retrieved documents \cite{benedict2024gen}, making the standard retrieval systems still vital to satisfy our information needs. Therefore, building test collections and its risks continues to be an important area to explore.

The advancements in generative AI technologies led several researchers to focus on how LLMs can be utilized in the evaluation of IR systems \cite{rahmani2024report}. Several studies have employed these models to automatically assess document relevance \cite{faggioli2023perspectives,upadhyay2024llms,thomas2024large,rahmani2024syndl} and generate search queries \cite{rahmani2024synthetic,rajapakse2023improving} to construct test collections from crawled document sets. To the best of our knowledge, no prior research has attempted to use recent LLMs to generate   test collections entirely, including both documents and relevance judgments.

%In this study, we explore whether we can use LLMs to generate test collections including documents and relevance judgments to evaluate IR systems,  diverging from the conventional test collection construction approach significantly. In particular, we first gather search topics and create sub-topics for each using GPT3.5 to be able to cover different aspects of a given topic. Next, for each subtopic, we generate a document, assuming that the generated document will be relevant to the corresponding topic. As generated documents might be trivial to retrieve, we generate also non-relevant documents that are similar to relevant documents. Lastly, we generate documents in random topics to increase the collection size, assuming that these documents will not be relevant to any topic, similar to considering not-judged documents as non-relevant in conventional pooling method. In our approach, we generate all data except the first topics. However, we utilize existing search topics for fair comparison in our experiments. This process can be also avoided by applying topic generation methods, e.g., [REF].
In this study, we investigate whether LLMs can be utilized to generate test collections and introduce \testcol{} as a proof-of-concept test collection that is generated by an LLM. Our approach significantly diverges from the conventional test collection construction approach by generating documents. Our proposed approach relies on the intuition that a generated document should be relevant to its prompt. Thus, if we use  search topics as prompts and generate documents accordingly, we do not need to collect relevance judgments because we can decide the relevance of documents during the generation phase.  
Specifically, we first collect existing search topics from TREC tracks and generate subtopics for each of them using GPT 3.5, to increase topical diversity.  
For each subtopic, we generate a document, assuming that the generated document is relevant to the corresponding topic. Considering that generated documents may be too easily retrieved, we also generate non-relevant documents that are similar  to the relevant ones. Additionally, to further enlarge the collection, we generate documents on random topics, assuming that these documents will not be relevant to any of the search topics, similar to the conventional pooling method where unjudged documents are treated as non-relevant. Eventually, \testcol{} consists of 96,196 documents, 300 topics, and 18,964 relevant judgments, costing only \$ 126 based on the current charges. 

We conduct a comprehensive set of experiments to analyze the quality of the generated documents, the accuracy of the relevance judgments, and the reliability of system evaluations done by \testcol{}. Specifically, we compare (i) the generated documents with the Disks 4-5 document collection used in several TREC tracks \cite{trec8-overview,eguchi2002overview,harman1996overview}, and (ii) the system ranking correlation between TREC collections and \testcol{}. Our observations are as follows. i) \testcol{}  contains shorter documents and sentences, with lower lexical diversity but require a higher education level (based on established readability metrics),  compared to human-authored
documents. ii) The generated documents assumed to be relevant are indeed relevant in 83\% of the cases. However, the accuracy of the relevance ``judgments" varies across topics such that  almost all generated documents are relevant in several topics. On the other hand, the generated documents are related to the search topic but not addressing the actual information need in some search topics. %So our approach will be more effective if we select topics carefully, paying attention to this concern. 
iii) The ranking of IR systems based on \testcol{} is highly similar to the ranking of systems based on the existing TREC tracks for P@100, reaching 1.0 and 0.95 Kendall's $\tau$ score for TREC6 and Robust2004, respectively. While the ranking correlation with MAP and RPrec metrics do not reach 0.9 threshold, which is a conventionally accepted threshold for high ranking correlation \cite{voorhees2000variations}, we find that  the ranking correlation across existing TREC collections usually do not reach this threshold and our results are compatible with them. However, our system rankings are highly different than TREC rankings based on P@10.

%Although most of the data in our approach is generated, we use existing search topics for a fair comparison in our experiments, though this could be bypassed by employing topic generation methods such as [REF].

%During our study, we present a comprehensive comparison between our collection and the traditionally create TREC collections, in terms of many aspects such as readability, lexical diversity, topical diversity, systems rankings. Our findings show that artificial test collections might be alternative for traditional collections. 

\begin{comment}
The main contributions of our work are as follows.
\begin{itemize}
    \item  We explore how to generate a test collection using LLMs based on  only search topics.  Our proposed method allows researchers to design the document collection and the number of relevant documents based on their needs. %we can decide relevance of documents while generating them, which resolves the excessive cost of annotating documents manually.
    \item We introduce \testcol{} which is the first  test collection with generated documents (and relevance judgments) using GPT3.5
    \item In our experiments, we extensively evaluate \testcol{}  and find that \hl{blah blah blah}
    \item We share our code and data to increase the reproducibility of our paper and provide an important data resource for the research community to test and train their IR systems\footnote{The URL is hidden due to the double blind review process}.
\end{itemize}
\end{comment}

The main contributions of our work are as follows.
\begin{itemize}
    \item  We explore the potential of using LLMs to generate a test collection. Our proposed method enables researchers to design document collections and configure the number of relevant documents and their topics based on their specific needs.
    \item We introduce \testcol{}, the first  test collection, consisting of LLM-generated documents and corresponding relevance ``judgments".
    \item Through extensive experiments, we evaluate the effectiveness of \testcol{} and demonstrate that it yields compatible rankings with TREC collections for P@100, MAP, and RPrec metrics.
    \item We share our code and data to enable reproducibility of our results and to support the research community in testing and training their IR systems.%\footnote{The URL will be shared when the paper is published}.
\end{itemize}

The rest of the paper is organized as follows. In Section \ref{sec:related_work}, we review previous work related to our study. In Section \ref{sec:automatic_document_generation}, we define our methodology for generating a test collection using an LLM.
We introduce \testcol{} in Section \ref{sec:chatgpt23_analysis} and provide a quantitative analysis of the generated documents in Section \ref{sec_quan}.
%\textbf{Section \ref{sec:chatgpt23_analysis}} presents an analysis of the properties and characteristics of the generated collection. 
We present our experimental results for system rankings with \testcol{} in Section \ref{sec:evaluation}.
 We discuss the limitations of our work in Section \ref{sec:discussion_limitations} and conclude in Section \ref{sec:conclusion}.

%The rest of the paper is organized as follows. We first discuss the related studies in Section \ref{sec_rel_work}. In Section \ref{sec}, we explain our method to generate a test collection using an LLM.  We present our experimental results for system ranking with \testcol{} in Section \ref{}. We discuss the limitations of our work in Section \ref{} and conclude in Section \ref{}.

%RQ-1: bir konuya farklı açılardan belgeler nasıl oluşturulur

%RQ-2 aynı konuda farklı yazım stilleri elde edebiliyor muyuz?

%Rq-3: ilgisiz belge nasıl elde ederiz.

%rq-4 oluşturulan test koleksiyonu işe yarıyor mu? Belki de sadece data augmentation için kullanılabilir.

\section{Related Work} \label{sec:related_work}

%In this section, we provide a comprehensive discussion of studies about IR evaluation. Firstly, we discuss the studies that build test collection with human annotations (Section \ref{sec_rel_human}). Next, we present pseudo-test collections which require no relevance judgments (Section \ref{sec_rel_preudo}). Subsequently, we explain studies for simulated test collections that contain generated queries and/or documents (Section \ref{sec_rel_sim}). Lastly, we discuss the studies that use modern LLM models for IR evaluation (Section \ref{sec_rel_llm}). 
This section presents a detailed examination of research on IR evaluation methodologies. We begin by discussing studies that focus on the construction of test collections using human annotations (Section \ref{sec_rel_human}). Next we discuss the pseudo-test collections, which eliminate the need for relevance judgments (Section \ref{sec_rel_preudo}). We then explore simulated test collections that generate queries and/or documents (Section \ref{sec_rel_sim}). Finally, we review recent research studies utilizing LLMs for IR evaluation (Section \ref{sec_rel_llm}).

%bu makalenin ilfgili olduğu alanlar

\subsection{Test Collection Construction with Human Annotations} \label{sec_rel_human}

%IR community has shown great interest in shared-tasks such as TREC\footnote{https://trec.nist.gov/}, NTCIR\footnote{https://research.nii.ac.jp/ntcir/index-en.html}, FIRE\footnote{http://fire.irsi.org.in/fire/2024/home}, and CLEF\footnote{https://www.clef-initiative.eu}, in which several test collections have been constructed for various IR tasks \cite{eguchi2002overview,wt14,kanoulas2018clef}. In the literature, there also exists test collections that have been developed without running a shared-task. Hasanain et al. \cite{hasanain2020artest} use interactive search \cite{cormack1998efficient} to identify documents to be judged instead of applying pooling method \cite{jones1975report} based on system results. Rahman et al. \cite{rahman2020efficient} explore employing active learning methods to build test collection without any shared task. In all these studies, test collections are built using crawled human-authored documents while we generate documents using LLMs and do not require any shared-task nor relevance judgment.

The construction of test collections has attracted considerable interest due to its critical role in evaluating IR systems and the inherent challenges it presents. In particular, the process is resource-intensive due to the substantial amount of judgments required. Therefore, optimizing evaluation budgets is a key concern in achieving reliable system evaluations. A significant amount of research has focused on selecting documents for judgment, given that it is infeasible to assess every document-topic pair within a collection. The pooling method \cite{jones1975report}, which focuses on evaluating the top-ranked documents from pre-selected IR systems, has been widely adopted by the IR community  \cite{moghadasi2013low}. While pooling reduces costs and can produce high-quality test collections (e.g., TREC-8 \cite{trec8-overview}), several challenges remain, including incomplete judgments, the necessity for multiple IR systems, and the high cost of relevance judgments. Consequently, numerous studies have tried to refine pooling through various strategies, such as predictive evaluation metrics for incomplete judgments \cite{buckley2004retrieval,sakai2007alternatives}, smart document selection methods \cite{carterette2005incremental,moffat2007strategic,pavlu2007practical}, and relevance inference using classifiers \cite{buttcher2007reliable,aslam2007inferring,macavaney2023one}. Other approaches have included crowdsourcing \cite{grady2010crowdsourcing,mcdonnell2016relevant,cormack2018beyond} and reducing the number of topics \cite{kutlu2018intelligent,roitero2018effectiveness,guiver2009few,hosseini2012uncertainty}.

The IR community has also shown great interest in shared-tasks such as TREC\footnote{https://trec.nist.gov/}, NTCIR\footnote{https://research.nii.ac.jp/ntcir/index-en.html}, FIRE\footnote{http://fire.irsi.org.in/fire/2024/home}, and CLEF\footnote{https://www.clef-initiative.eu}, which have facilitated the construction of several test collections for various IR tasks \cite{eguchi2002overview,wt14,kanoulas2018clef}. In addition to shared-task test collections, some studies have developed collections independently. For instance, Hasanain et al. \cite{hasanain2020artest} utilize interactive search \cite{cormack1998efficient} to identify documents for relevance judgments rather than relying on traditional pooling methods \cite{jones1975report}. Similarly, Rahman et al. \cite{rahman2020efficient} explore active learning techniques to build test collections without a shared-task setup. 

All these approaches rely on human-authored documents; in contrast, our work generates documents using LLMs and eliminates the need for both organizing a shared task and collecting relevance judgments.

\subsection{Pseudo Test Collections} \label{sec_rel_preudo}

%Due to the high cost of human annotations, several researchers explored alternative resources that enable constructing test collections without any annotation. A number of researchers focused on predicting the relevance of documents based on participants of shared-tasks.  For instance, Nuray et al. \cite{nuray2006automatic} investigate merging retrieval results of multiple IR systems to form \textit{pseudo relevant} documents and evaluate systems accordingly. Hauff and de Jong \cite{hauff2010retrieval} report that evaluation with no judgment correlates with evaluation with few judgments. However, Roitero et al. \cite{roitero2020effectiveness} demonstrate that straightforward combination strategies utilizing data fusion techniques are generally ineffective and may even be harmful.
Due to the high cost of human annotations, researchers have investigated alternative approaches to constructing test collections without manual annotations. One such approach involves predicting document relevance based on shared-task participants. For example, Nuray et al. \cite{nuray2006automatic} explore the combination of retrieval results from multiple IR systems to form pseudo-relevant documents for system evaluation. Similarly, Hauff and de Jong \cite{hauff2010retrieval} report that evaluation without judgments correlates with evaluation using few judgments, although Roitero et al. \cite{roitero2020effectiveness} highlight the limitations of simple combination strategies based on data fusion techniques.

As another research direction, prior work investigated how particular data resources can be turned into a test collection automatically, including web pages \cite{asadi2011pseudo}, tweets \cite{berendsen2013pseudo}, Wikipedia dumps \cite{dietz2022wikimarks}, and digital libraries \cite{berendsen2012generating}. For instance, Asadi et al. \cite{asadi2011pseudo}   propose using anchor texts in web pages as queries and assume documents pointed by anchor texts are relevant to the corresponding query. This enables them to construct a test collection without any manual effort to create topics and collect judgments. While they do not use their method to evaluate systems,  they show that it can be utilized to train learning-to-rank models.

%Another line of research has focused on automatically transforming particular data resources into test collections, including web pages \cite{asadi2011pseudo}, tweets \cite{berendsen2013pseudo}, Wikipedia dumps \cite{dietz2022wikimarks}, and digital libraries \cite{berendsen2012generating}. For example, Asadi et al.  \cite{asadi2011pseudo} propose using anchor texts from web pages as queries, assuming that the linked documents are relevant to the query, thereby enabling the creation of test collections without manual effort in generating topics or collecting judgments. While their method is not applied for system evaluation, it has been shown to be useful for training learning-to-rank models.

%Similarly, Dietz et al. \cite{dietz2022wikimarks} construct queries from the titles and subheadings of Wikipedia pages and assume that the text following each title or heading is relevant to its respective query, eliminating the need for manual human judgments. Berendsen et al. \cite{berendsen2013pseudo} assume that tweets are relevant to the hashtags used in the corresponding tweets. Therefore, they generate queries from hashtags and build the test collection accordingly for training and tuning LTR systems for microblog retrieval. Dietz and Dalton \cite{dietz2020humans} build a test collection for passage retrieval using queries derived from pages in Wikipedia, textbook chapters, product descriptions, and others. Only passages from the document that a particular query is generated are considered relevant to that query.
Similarly, Dietz et al. \cite{dietz2022wikimarks} generate queries from the titles and subheadings of Wikipedia pages, assuming that the text following each title or heading is relevant to the corresponding query, thereby eliminating the need for manual relevance judgments. Berendsen et al. \cite{berendsen2013pseudo} take a similar approach with tweets, assuming that tweets are relevant to the hashtags they contain. They generate queries from hashtags and build a test collection to train and fine-tune learning-to-rank systems for microblog retrieval. Furthermore, Dietz and Dalton \cite{dietz2020humans} develop a test collection for passage retrieval, using queries derived from Wikipedia pages, textbook chapters, and product descriptions. In this approach, only passages from the documents in which the queries were generated are considered relevant.

%Nevertheless, in all these studies, an existing document collection is required while in our approach we also create our own documents.
Nevertheless, while all these studies rely on pre-existing document collections, our approach generates the documents, thereby eliminating the dependency on existing corpora.

%\cite{berendsen2012generating} proposes a method for creating pseudo test collections in digital libraries, where data is limited but well-annotated. By leveraging these annotations and associated documents, we generate query-document pairs. We explore how different sampling methods for these annotations affect learning-to-rank performance and compare the results with traditional editorial topics and judgments. Our findings show that training a learning-to-rank algorithm using these pseudo judgments is feasible.

%\cite{dietz2020humans} build a test collection for entity-passage retrieval that does not require any human assessor. Our approach relies on a collection of a humancreated corpus, which is often readily available, such as a Wikipedia dump, textbook chapters, product descriptions, a knowledge compendium, or glossary. Queries are derived from the titles of input pages, which we refer to as title in the following. In our example domain, the information need is interpreted as “Provide comprehensive information about title”.

\subsection{Simulated Test Collections}\label{sec_rel_sim}

%Our test collection can be considered as a simulated test collection. While we observe remarkable achievements in text generation recently after decades of research, research on simulated test collections have started decades ago \cite{hawking2020simulating}. In particular, the first simulated test collection has been developed by Cooper \cite{cooper1973simulation} in 1973. Cooper created an artificial text block by generating words based on their frequency distribution.  Next he created a thesaurus based on the word associations in his generated text and used the  thesaurus to generate documents and queries. In total he created 150 documents for his experiments. %Griffiths \cite{} have extended his work by generating queries and relevance judgments.
Our test collection falls under the category of simulated test collections. Research on simulated test collections has a history spanning several decades \cite{hawking2020simulating}. Notably, the first simulated test collection was introduced by Cooper in 1973 \cite{cooper1973simulation}.  In his work, Cooper created artificial text blocks by generating words based on their frequency distributions. He then created a thesaurus from the word associations in this generated text, using it to construct both documents and queries. For his experiments, Cooper generated a total of 150 documents.

%As a noteworthy work, Zobel and Moffat \cite{zobel1998inverted} created a simulated test collection using a language model which generates documents with similar statistical features of real text. They also generate queries using the vocabulary of the corpus. However, they do not use their collection to rank systems, but to compare document signatures and inverted files.
As a noteworthy work, Zobel and Moffat \cite{zobel1998inverted} developed a simulated test collection using a language model to generate documents with statistical properties resembling real text. In addition, they generated queries based on the vocabulary of the corpus. However, their focus was not on ranking information retrieval systems, but rather on comparing document signatures and inverted file structures.

%As we already know which documents are relevant to a given topic, our approach might be considered conceptually similar to studies for building a simulated test collection for known-item search \cite{azzopardi2007building,azzopardi2006automatic}. Similar to collection creation methods for known-item search, Kim and Croft \cite{kim2009retrieval} proposes a method to create a simulated collection for desktop research.   In these studies,  queries are generated for a given document, i.e., known-item, eliminating the need of collecting relevance judgments. However, we generate documents for a given topic, not the other way around. More importantly, our target task is ad hoc search, not known-item search.
Our method shares conceptual similarities with studies on known-item search \cite{azzopardi2007building,azzopardi2006automatic}, where relevance judgments are not required as queries are generated for specific documents. For instance, Kim and Croft \cite{kim2009retrieval} create a simulated collection for desktop research, where queries are generated for a specific document (i.e., known-item), thereby eliminating the need to collect relevance judgments. However, in contrast to these methods, we generate documents for a given topic rather than queries for a known-item. Furthermore, our focus is on ad hoc search tasks, not known-item search.

To our knowledge, the most comprehensive work on simulated test collections in the literature is done by Hawking et al. \cite{hawking2020simulating}. In their study, they develop several tools to extract features from a given corpus and generate synthetic corpora based on specific parameters and models. They explore various techniques, including unigram models, Markov models, and fine-tuned GPT-2 models. They implement three systems and evaluate their performance using the simulated test collection. While they find that the text generated by GPT-2 appears superficially plausible, it lacks meaningful content, making it unsuitable for queries that require understanding complex linguistic structures. In contrast, our work utilizes ChatGPT to generate documents and evaluates the resulting collection with a broader range of IR systems, including more recent retrieval technologies.

%SynthaCorpus implements a version of Azzopardi et al.’s best-performingmethod for generating known-item queries.v"
%In text generation, they explore unigram based language models, markov models, and encryption methods.  They use three systems and evalutate their performance on their test collection.
%They also fine-tune GPT-2 model with the TREC-AP collection. To ensure GPT-2 is comparable
%with the TREC-AP collection we generate the same number of documents (242,892) with the average document length (465 words per document).

%"Although GPT-2 generates superficially plausible text, the result has no meaning. This means that it can’t be used to support queries whose evaluation relies on advanced linguistic constructs."

\subsection{LLMs for Evaluation} \label{sec_rel_llm}

%With the remarkable success of LLMs in several tasks, researchers in IR community explored whether LLMs can be utilized to evaluate IR systems. Majority of these studies focus on using LLMs to collect relevance judgments \cite{faggioli2023perspectives,upadhyay2024llms,thomas2024large,rahmani2024syndl} investigating various prompts and strategies to improve their accuracy. LLMs have been also used relevance judging in specific tasks such as product search \cite{mehrdad2024large} and languages such as Tetun \cite{de2024exploring} and Brazilian Portuguese \cite{bueno2024quati}. In order to achieve a more reliable evaluation with LLM-judged document-topic pairs, \cite{oosterhuis2024reliable} develop methods to  create confidence intervals around ranking metrics derived from LLM-based relevance judgments. However, none of these studies use LLMs to generate documents and derive judgments accordingly. 
%These studies focus on various tasks such as product search \cite{mehrdad2024large}, 
With the remarkable success of LLMs across various tasks, researchers in the IR community have investigated whether LLMs can be employed to evaluate IR systems. Most of these studies have concentrated on using LLMs to collect relevance judgments \cite{faggioli2023perspectives,upadhyay2024llms,thomas2024large,rahmani2024syndl}, exploring different prompts and strategies to enhance their accuracy. LLMs have also been employed for relevance judging in specific tasks such as product search \cite{mehrdad2024large} and in less commonly studied languages like Tetun \cite{de2024exploring} and Brazilian Portuguese \cite{bueno2024quati}. To improve the reliability of evaluations involving LLM-judged document-topic pairs, some researchers \cite{oosterhuis2024reliable} have developed methods to create confidence intervals around ranking metrics derived from LLM-based relevance judgments. However, none of these studies utilize LLMs to both generate documents and derive relevance judgments, as we do in our study.

As generative IR becomes an emerging research direction with the advancements in LLMs, other researchers have focused on methodologies for evaluating generative IR systems \cite{sander2021exam,farzi2024exam}. However, our work centers on evaluating retrieval performance rather than generation capabilities.

Several researchers have also employed LLMs to generate synthetic test collections. For instance, Rahmani et al. \cite{rahmani2024synthetic} generated synthetic queries and relevance judgments using LLMs, using the MS MARCO v2 corpus as the document collection. Their findings indicate that synthetic test collections can produce evaluation results comparable to those derived from traditional test collections. Zhang et al. \cite{zhang2024usimagent} introduced USimAgent, a simulator that uses LLMs to mimic user search behaviors such as querying, clicking, and stopping. They report that USimAgent outperforms other methods in query generation and is as effective as traditional approaches in predicting user clicks and session termination. Similarly, Rajapakse and de Rijke \cite{rajapakse2023improving} generated queries using sequence-to-sequence models, demonstrating that dense retrievers can benefit from automatically generated training datasets. In contrast, our work does not focus on query generation but rather on generating document collections and relevance judgments.

%The development of synthetic datasets for training conversational agents has included new methods that use LLMs. The method \cite{abdullin2024synthetic} introduced a technique for generating synthetic dialogue datasets to help create linear programming models. This research is notable for demonstrating the flexibility of LLMs in producing datasets that are rich in context and tailored to specific goals, useful for developing and training goal-oriented conversational agents.

%Studies that generate documents can be considered  similar  to ours. Several researchers focused on utilizing LLMs to generate training data \cite{yu2024large,ye2022zerogen,yu2023regen}, particularly for text classification. However, we focus on search tasks. Regarding studies that work on generating documents for IR systems, Askari et al. \cite{askari2023test, askari2023expand} generate documents for given queries to augment training data. In our work, we explore whether these documents can be also used to evaluate systems, not only to enhance the training.   
Studies focused on document generation share similarities with our approach. Several researchers have explored the use of LLMs to generate training data, particularly for  text classification \cite{yu2024large,ye2022zerogen,yu2023regen}. However, we focus on search tasks. In the context of generating documents for IR systems, Askari et al. \cite{askari2023test, askari2023expand} generate documents for specific queries to augment training datasets. In contrast, our work investigates whether  generated documents can also be employed to evaluate systems, not just to enhance training processes.

\section{Automatic Test Collection Generation} \label{sec:automatic_document_generation}

%Cranfield-style test koleksiyonları çoktan var olan çok sayıda dokümanı derleme ve bu dokümanları çeşitli stratejiler doğrultusunda etiketleme sonucu oluşturulur. Bu çalışmada, bu yaklaşımı terkediyoruz ve dokümanları derlemek yerine onları üretmeyi tercih ediyoruz. Böylelikle bir dokümanın hangi konularla ilgili olduğu üretim aşamasında biliniyor ve etiketleme yükü ortadan kalkmış oluyor. Öte taraftan, manual olarak doküman üretmenin getireceği maliyet, çok daha büyük olabilir. Dolayısıyla, dokümanları otomatize bir şekilde ürettirmek kaçınılmaz oluyor.

Cranfield-style test collections are created by compiling a large number of already existing documents, creating/obtaining search topics, and labeling the relevance of document-topic pairs \cite{sanderson2010test}. In this study, we abandon this approach and prefer to generate documents from scratch using LLMs rather than crawling existing ones. This approach enables us to identify the topic of a document at the generation stage, eliminating the burden of labeling. Furthermore, we can control the prevalence of relevant documents for each topic and the difficulty of the retrieval task by this approach. 
However, there are several challenges that need to be addressed to achieve a reliable evaluation of systems. 

\begin{figure*}[!htb]
\centering
\includegraphics[width = 1\textwidth]{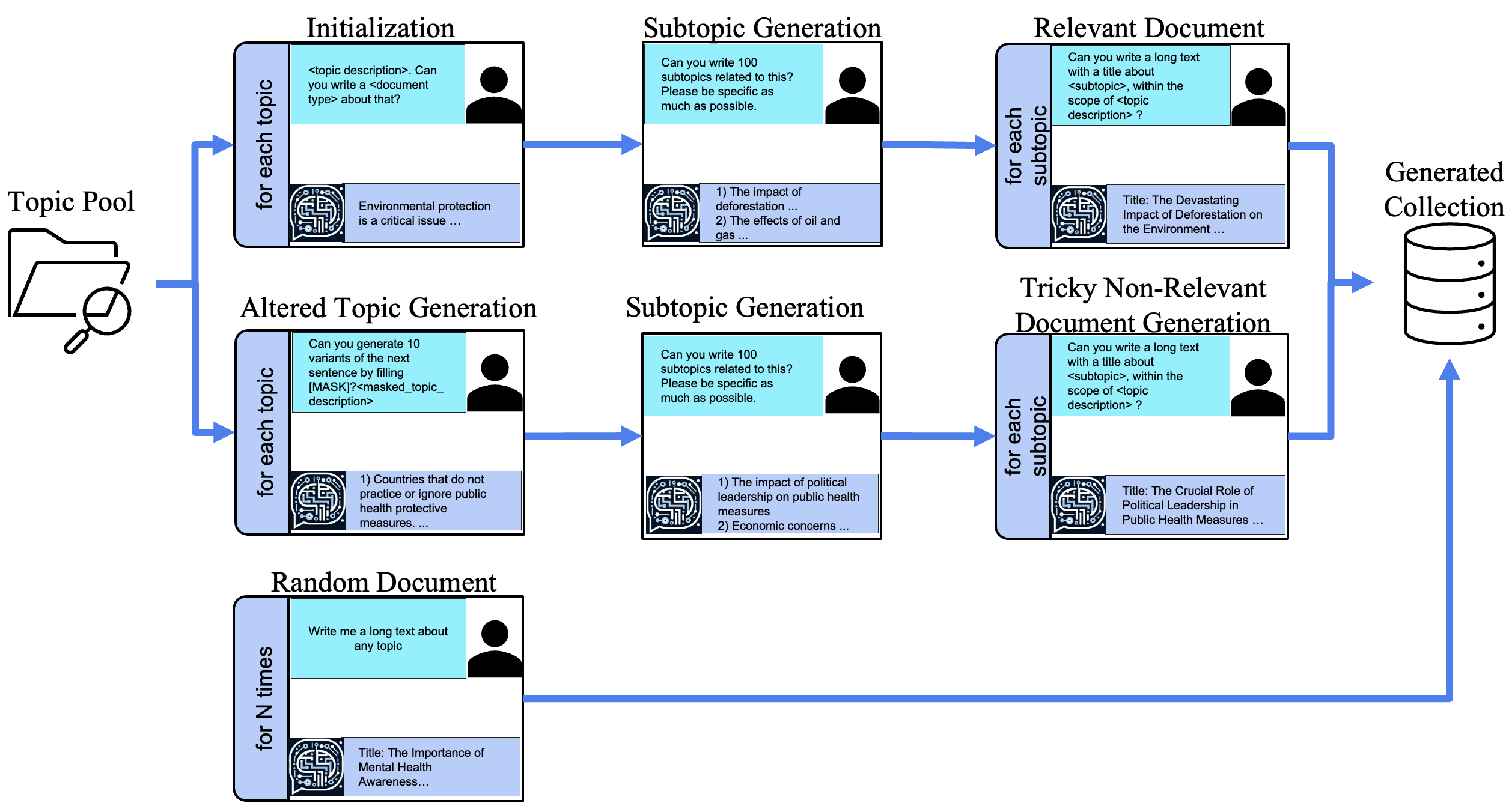}
\caption{Automatic Test Collection Generation Process and Prompts.}
\label{fig:process}
\end{figure*}

\textbf{Figure \ref{fig:process}} shows the overall process to generate a test collection. \textbf{Table \ref{tab_llm_prompts}} in Appendix provides the prompts and example outputs.  Firstly, we gather  a large set of search topics and generate a document using its topic description (i.e., initialization step). Next, in order to increase the diversity of documents, we generate several \textit{subtopics} for each topic using an LLM. Subsequently, we generate a document for each subtopic, assuming that the generated documents will be relevant to the corresponding topic. Furthermore, in order to make the retrieval task more challenging, we generate \textit{tricky non-relevant documents} which share similarities with relevant documents, but are eventually non-relevant. Lastly, to increase the collection size, we generate topics on any random topic, assuming that they will be non-relevant to any topic in our collection. Eventually, our test collection consists of search topics we gather and all generated documents. We do not collect any relevance judgments, but we consider that documents generated for a specific topic are relevant to the corresponding topic but non-relevant to the others. 
%we combine all generated documents to form our document collection. We use the gathered search topics for the evaluation of the systems with the corresponding relevant
%for each topic, we generate sub-topics using an LLM to generate various prompts and increase 

Now, we explain the details of %On the other hand, the cost of manually producing documents can be much greater. Therefore, it is inevitable to produce documents automatically.
%ChatGPT, insanların sorularına verdiği tatmin edici cevaplarla, piyasaya çıktığı günden itibaren ilgi odağı oldu. Bu doğrultuda, istediğimiz konularda dokümanlar üretmek için ChatGPT'yi kullanmaya karar verdik.
%ChatGPT has been the center of attention since  it was released in 30$^{th}$ November 2022, with its incredible success in generating high-quality texts. Thus  we use %Accordingly, we decided to use 
%ChatGPT to generate documents. % on the topics we wanted.
%Now, 
   generating subtopics (Section \ref{sec_increase_diversity}), documents relevant to a specific topic (Section \ref{sec_generate_relevant_documents}),  and  non-relevant documents. %which might be confused with relevant documents to make the retrieval task more challenging
(Section \ref{sec_generate_nonrelevant_docs}). 
%\todo{tüm bu processi/iş akışını anlatan bir figüre ihtiyacımız var}

%Prompts used to generate the collection are present at \textbf{Table \ref{tab_llm_prompts}}.

%\textbf{Table \ref{tab_llm_prompts}} presents samples documents in our collection.
%\hl{yorum yorum yorum}

%For this study, we need a topic set, to which generated documents will be considered relevant or non-relevant. To this end, we use topics of TREC-8 collection. In this way, we will be able to compare our collection to its conventionally created version. In other words, how system scores are affected when a synthetic collection including the exact same topics can be observed.

%\subsection{How to generate text for subtopics}
%\subsection{How to generate relevant documents: A simplistic method}

\subsection{Generating Subtopics} \label{sec_increase_diversity}

Using LLMs, we can generate several relevant documents for each topic at a minimal cost. However, having highly similar relevant documents would reduce the reliability of system evaluation, as retrieving one document may make it easier for the system to retrieve others that are closely related.
%In this way, a relevant document can be produced for each topic, but one relevant document per topic 
%would not be sufficient, of course.
%\subsection{Augmenting Relevant Documents}
Therefore, we need a diverse set of relevant documents for each topic to construct a realistic and reliable test collection. 

%While generating a single relevant document is easy, it is challenging to create a diverse set of  relevant documents because we need various relevant documents for each topic to construct a realistic test collection. 

As a naive approach to creating different documents, we can ask the LLM to regenerate its response for a given prompt (i.e., topic) several times. However, using the same prompt usually yields highly similar documents. As another approach, we can change the prompt slightly for each generated document. For instance, in our not-reported experiments with ChatGPT, we explored 
 using follow-up prompts such as "Talk about it more" and "What else can you say?" to create different documents. However, this method also failed and created documents with high overlap compared to the other generated documents we generated. 
%tried different prompts to create different but relevant documents for a single document.
%over and over again was the first thing we tried, in order to produce multiple relevant documents per topic. We accepted each response as another relevant document. In addition, we tried queries such as "Talk about it more", "What else can you say?". However, this method failed to generate distinct documents. Although documents generated in this way keep their relevancy, they all cover the same information. Also, the overlap rate of words is quite high. 
%. This situation will lead to poor evaluation of systems. 

In order to address this issue, we take the following steps. Many topics are broad and can be addressed in different ways. For instance, if a topic asks about technological developments in $20^{th}$ century, this information need can be addressed from various perspectives, such as technological developments in medicine, transportation, and other domains.
%several answers from different perspectives can be given for this information need such as technological developments in medicine, technological developments in transportation, and others. 
Thus, instead of using only the original topic description to generate documents, we first use an LLM to produce subtopics related to the given topic, and then generate documents for each subtopic. Because we use distinct prompts for each document, the resulting documents likely exhibit less word overlap compared to those generated based on only the original topic description.  To generate subtopics for a given topic, we provide the LLM with the topic description, followed by the prompt: ``\textit{Can you write 100 subtopics related to this? Please be as specific as possible.}". \textbf{Table \ref{tab_subtopic_prompts}} lists the first 10 subtopics generated for topic 260 of TREC-5.

\begin{table}[!htb]
\small
\centering
\renewcommand{\arraystretch}{1.2}
  \begin{tabular}{  r  | l }
    \hline
      &  \textbf{Generated Subtopics} \\ \hline \hline
  1 & The significance of the Neolithic era in human history \\ \hline
2& How scientists determine the age of ancient artifacts and fossils \\ \hline
3& The origins of agriculture and domestication of animals \\ \hline
4& The impact of climate change on the development of human civilization \\ \hline
5& The spread of human populations across the globe \\ \hline
6& The role of women in ancient societies \\ \hline
7& The evolution of human technology during the Neolithic era \\ \hline
8& The development of writing systems and communication \\ \hline
9& The use of fire and its importance to human survival \\ \hline
10& The emergence of social hierarchy and class systems 
 \\ \hline
  \end{tabular}
  \caption{The first 10 subtopics generated by ChatGPT for the topic 260 of TREC-5, of which the topic description is ``Evidence of the existence of human life 10,000 years ago.''} 
  \label{tab_subtopic_prompts}
\end{table}

%we attack the problem from two different perspectives: i) diversity in content and ii) diversity in writing style. In particular, many topics are broad and can be addressed in different ways. 
%we first use an LLM to generate sub-topics related to the given topic description (\textit{see ``Subtopic generation'' prompt at Table \ref{tab_llm_prompts}}).
%Subtopics can be described as smaller units that cover different aspects of a topic. 

\subsection{Generating Relevant Documents}\label{sec_generate_relevant_documents}

We generate relevant documents using the original topic description and subtopics derived from it. However, in order to identify the most effective method, 
%\hl{topic description and also separate documents for each subtopic}.
%Before generating the whole collection, 
we conducted a pilot study %using five random topics from TREC-8 (402, 416, 419, 429, 438) 
and examined the impact of prompt memorization and the specification of document type on the generated documents.

\noindent
\textbf{Memorization of Previous Prompts.} Instruct-tuned LLMs, e.g., ChatGPT,  keep the previous prompts and generate texts considering all prompts to achieve coherent conservation. This might negatively impact the textual diversity as the model tends to shape its all responses around the very first request. As expected,  in our pilot study, we %created 100 subtopics for  Topic 402 of TREC-8 and then 
%created news articles for each subtopic using ChatGPT. When we allow memorization, we 
observed that the minimum, maximum, and median textual similarity\footnote{We used ``SequenceMatcher'' module from Python ``difflib'' library to calculate the similarity between two texts.} between generated document pairs is 0.006, 0.977, and 0.125, respectively. However, if we start a new chat session for each generated document (i.e., no memorization), the minimum, maximum, and median textual overlap between generated document pairs reduce to 0.004, 0.202, and 0.055, respectively. Therefore, 
%One of the problems we encountered during collection creation is generated documents' being highly similar. To exemplify, documents generated for different subtopics of the same topic were containing a lot of common sentences and patterns. The key reason for this problem is actually ChatGPT's chat session memory. ChatGPT is able to remember what you have asked before in the same chat session and form its responses based on chatting history. 
 we do not use any memorization in our document generation processes. %to increase the diversity of the generated documents.
%if started a chat by asking ChatGPT to generate a document about a subtopic, ChatGPT tends to shape its following responses around the very first request. As a result, documents generated in the same chat session turn out to be similar in terms of content and writing style. 
%To overcome this situation, we tried restarting the session before each generation process. In this way, we eliminated any kind of dependence between different documents. According to our observations, obtained documents were more diverse in terms of content and writing style.

\noindent
\textbf{Document Type.} In our pilot study, we generated news articles as mentioned above. However, specifying a particular writing style might reduce the textual diversity of the collection. 
% restriction might cause t the similarity of the documents. 
%While the generated documents  are indeed mimicking news articles (e.g., mentioning the words of experts and scientific researches carried out), we observe that this writing style restriction increases the similarity of the documents. 
Therefore, we also asked GPT3.5 to generate just long texts without mentioning the document type and calculated their textual overlap. %(\textit{see ``Document generation'' prompt at Table \ref{tab_llm_prompts}}), yielding longer informative documents with no specific style.
%One of the key features of ChatGPT is that it can return a response in different formats if desired. Thinking TREC-8 consists of news articles, we first generated documents in news article format. Obtained documents were indeed mimicking a news article by mentioning the words of experts, scientific researches carried out, etc. In addition to news articles, we asked ChatGPT to generate long texts. The resulting documents were longer informative documents.
%For Topic 402, 
We observed that the minimum, maximum, and median textual overlap between \textit{long texts} is 0.001, 0.113 and 0,018, respectively. As 
%Comparing generated news articles vs. long texts, we observe that %When we measured the word overlap, we observed that 
the diversity of  words is higher in long texts than news articles,  %since ChatGPT uses similar patterns, phrases to keep the text in news article format. 
%Because of fewer word overlap and longer textual volume, we decided to 
we opt for creating our dataset with  long texts rather than news articles.

\subsection{Generating  Non-Relevant Documents}\label{sec_generate_nonrelevant_docs}

%Doğru çalışan bir değerlendirme için, koleksiyondaki ilgili belgeler kadar ilgisiz belgeler de kritik bir öneme sahiptir. Örneğin, bütün dokümanların bütün topiclerle ilgili olduğu bir koleksiyon ile test edilen bütün IR sistemleri en yüksek puanı alacaktır. Ancak koleksiyondaki ilgililik oranı düştükçe, retrieval task'ı sistemler için daha zor bir hale gelir ve sadece ilgili dokümanları ilgisizlerden ayırt edebilenler puan alır. Dolayısıyla ilgilisiz koleksiyondaki ilgisiz dokümanların sayısı ve niteliği üstüne düşünmeye değer konulardan bir tanesi.

For a robust and reliable evaluation, the quantity and quality of non-relevant documents are equally as critical as those of relevant documents. For instance, if all non-relevant documents are entirely different than the relevant ones, retrieval systems may identify all relevant documents, limiting the ability to differentiate between systems based on their retrieval performance. In addition, as the proportion of relevant documents within the collection decreases, the retrieval task becomes  more challenging for the systems.  However, an excessively challenging evaluation setup may also limit our ability to accurately distinguish the performance of different retrieval systems. Thus, balancing these factors is essential to ensure a reliable evaluation. %, and only those who can distinguish relevant documents from irrelevant ones get points. Therefore, the number and quality of irrelevant documents in the collection is one of the issues worth thinking about.

%En basit yaklaşımlardan birisi, topiclerin kendi için üretilen dokümanları ilgili, diğer topicler için üretilen dokümanları ilgisiz olarak düşünmek olabilir. Bu aslında mantıklı bir yaklaşım olabilir, ancak bu durumda ilgisiz dokümanlar ile ilgililer birbirinde tamamen farklı olacaklardır. Dolayısıyla da, IR sistemleri için ilgileleri ayırt etmek oldukça kolaylaşacak ve koleksiyonun ölçücülüğü azalacaktır. Bu sorunu adreslemek için, bahsedilen ilgisiz dokümanlara ek olarak, tricky non-relevant dokümanlar oluşturuyoruz. Bu dokümanlar aslında söz konusu topic'te belirtilen bilgi ihtiyacını karşılamıyor ancak bu topic'le benzer konularda oluşturulmuş dokümanlar. Yani sırdan bir ilgisiz dokümana göre, ilgili bir dokümanla kullanılan kelimeler ve konsept açısından daha çok ortak yönleri bulunuyor. Bu yüzden, değerlendirilen sistemlerin bu dokümanları ilgilerden ayırt etmesi çok daha zor olabiliyor. 

In standard test collections, unjudged documents are typically assumed to be non-relevant. Inspired by this assumption, documents generated for a specific topic can be regarded as relevant to that topic, while those generated for other topics are treated as non-relevant. While this is a reasonable starting point, these relevant and non-relevant documents may be about entirely different topics, potentially making the retrieval task too easy. Therefore, in addition to these non-relevant assumed documents, we also create two additional types of non-relevant documents: \textit{tricky} non-relevant documents and documents on random topics.

\noindent
\textbf{Tricky non-relevant documents.} In this step, our objective is to generate documents that are challenging to distinguish from relevant ones. These documents should exhibit similarities to relevant documents in terms of language and concepts but ultimately fail to satisfy the specified information need. 
%We generate documents that are harder to distinguish from the relevant ones, thereby making the retrieval task more challenging. These documents do not actually meet the original information need but are relevant to a similar topic. Thus,  these documents will have more in common with a relevant document in terms of words and the concept compared to an ordinary non-relevant document. %Therefore, it can be much more difficult for evaluated systems to distinguish these documents from interests.
%Bu tarz ilgisiz dokümanlar oluşturmak için yine ChatGPT'nin yardımına başvurduk. Öncelikle bir topic'in description cümlesindeki çeşitli segmentleri maskladık. Masklayacağımız kelimeleri seçerken, topic'in title'ında yer alan keyword'leri kullandık. Masklanan description'ı ChatGPT'ye vererek, [MASK] bulunan yeri doldurmasını istedik. 
To create such documents, we first mask various words in a given topic's description. In this masking process, we employ a named entity recognition model to identify keywords in the topic description and replace them with the ``[MASK]'' token. Next, we prompt GPT-3.5 to create 10 variants of the topic description by filling ``[MASK]'' tokens\footnote{The prompt is as follows. "Can you generate 10 variants of the next sentence by filling [MASK]: $<$masked topic description$>$ -- Example: $<$topic description$>$"}. Subsequently, we apply the same procedure used to generate relevant documents. In particular,  for each of the altered topic descriptions, we create five relevant documents, yielding a total of 50 non-relevant documents for the original topic description.
%Subsequently, we apply the same procedure used for relevant document generation to create five relevant documents to each altered version of the topic description, yielding 50 non-relevant documents to the original topic description.

\noindent
\textbf{Documents on random topics.} As test collections should contain a large set of documents, we also generate documents without specifying any particular topic and assume they are non-relevant to all topics. %Table \ref{tab_llm_prompts} provides the prompt used for these documents, along with an example output. 
There are two potential concerns regarding these documents. First, they might not be necessary if we utilize a larger number of search topics. 
However, identifying and curating such an extensive set of topics may not always be feasible. Second, the assumption that these documents are non-relevant to any topic might not always hold true. This concern, however, parallels the practice in standard test collections, where unjudged documents are typically treated as non-relevant. Nevertheless,  we evaluate the impact of these documents on system performance in our experiments (Section \ref{sec_exp_results}) to address these issues.

 %We gave the masked description to ChatGPT and asked it to fill in the [MASK].

%- subtopic'ler ile manuel olarak müdahale etme

%- assuming that documents generedy for other topics are non-relevant

\section{\testcol: The first LLM-generated test collection} \label{sec:chatgpt23_analysis}

Using our proposed approach, we created the very first collection using ChatGPT, namely, \testcol.  %We use ChatGPT for the all text generation processes.  %In order to automatize text generation process, we use pyChatGPT\footnote{https://github.com/terry3041/pyChatGPT} library. %Once ChatGPT+ is available, we signed up for the premium account to speed up the text generation process. 
We used ad hoc search topics of TREC-5 \cite{trec5}, TREC-6 \cite{voorhees2000overview}, TREC-7 \cite{trec7}, TREC-8 \cite{trec8}, and Robust2004 \cite{robust2004}, yielding 300 topics in total, as our basis to create relevant documents. 
%to compare and assess the reliability of system evaluation based on our test collection (See Section \ref{sec_exp}). Therefore, we use its search topics as our basis to create relevant documents. 
In particular, for each topic, we first generated a relevant document, 50 tricky non-relevant documents, and subtopics using the topic description. Subsequently, for each subtopic, we generated a \textit{relevant} document. We observed that ChatGPT occasionally fails to generate 100 subtopics for certain topics, resulting in a reduced number of relevant documents for those cases.  %We consider a document generated using a topic description relevant to that topic but non-relevant to others.
%In addition, for each topic, we generate  50 tricky non-relevant documents.  
Lastly, in order to increase the size of the collection, we created 59,804 documents  on random topics. % (\textit{see ``Subtopic generation'' prompt at Table \ref{tab_llm_prompts}}). We consider these documents non-relevant to any topic we use for evaluation.
Overall, we generated 96,196 documents (consisting of approximately 41.3M words).  Among these documents, 18,964 of them are considered  relevant to the respective topic.

%In this section, we investigate the cost of creating such collection by using ChatGPT in various aspects. Firstly, we focus on the financial aspect. Currently, ChatGPT API charges \$1.50 / 1M for input tokens and \$2.00 / 1M output tokens. Our prompts consists of 1,443,000 words in total. According to OpenAI's explanation, one token typically corresponds to 0.75 words on average for English. Considering this, we have 1,924,000 input tokens. The number of words generated, including the documents and the responses generated during intermediate steps (e.g. subtopic generation), is 46,272,943, which equals to 61,697,257 output tokens. Based on those numbers, the price for using API adds up to only \$126.
%Secondly, we calculate the energy cost. Although our approach for collection generation is free of human labor, it shouldn't be overlooked that LLMs consume energy. Therefore, we calculated the approximate energy consumed during the generation of \testcol{}. According to itself, ChatGPT-3.5 consumes roughly 0.3 watt-hours (Wh) per query for typical prompts of around 20 tokens on average, which means 0.015 Wh per token. We also found online articles confirming those numbers \footnote{https://www.baeldung.com/cs/chatgpt-large-language-models-power-consumption}. Again, according to ChatGPT, one token typically corresponds to 0.75 words on average for English. Considering \testcol{} contains 44,718,147 words, the total energy cost is approximately 954 kWh.
%-----------
To evaluate the feasibility and sustainability of using LLMs in test collection generation, we analyzed the costs associated with generating \testcol{} using GPT-3.5, focusing on both financial and energy consumption aspects.

%We analyze the costs associated with generating the collection using ChatGPT, focusing on both financial and energy consumption aspects.

\begin{itemize}
    \item \textbf{Financial Cost.} The ChatGPT API charges \$1.50 per million input tokens and \$2.00 per million output tokens. To create \testcol{}, we used approximately 1.924M tokens for prompts. 
%The total word count of our prompts amounts to 1,443,000 words. According to OpenAI’s documentation, one token corresponds to approximately 0.75 words on average for English. Based on this, the number of input tokens in our dataset is approximately 1,924,000. 
The generated texts including both documents and intermediate outputs (e.g., subtopic generation), are  61.7 million tokens in total. 
%61,697,257 output tokens. 
Given these values, the total price of using the API is approximately \$126.

\item \textbf{Energy Cost.} While our method for generating collections eliminates the need for human labor, it is essential to consider the energy consumption associated with using LLMs. We estimated the energy required to generate \testcol{}. Each query for a typical prompt of around 20 tokens consumes approximately 0.3 watt-hours (Wh), equating to 0.015 Wh per token\footnote{https://www.baeldung.com/cs/chatgpt-large-language-models-power-consumption}.
Given that we have approximately 63.6M tokens (i.e., input and output tokens) in total,
the corresponding energy consumption is estimated to be around 954 kWh.

\end{itemize}

\section{Quantitative Analysis }\label{sec_quan}

In this section, we quantitatively analyze our collection \testcol{} from several perspectives. In particular, we first compare \testcol{} against the documents of Disks 4-5 which have been used in previous TREC tasks \cite{harman1996overview,trec8} in terms of document structure (Section \ref{sec_ana_doc_str}), lexical diversity (Section \ref{sec:lexical_diversity}), readability (Section \ref{sec:readability}), and topical diversity (Section \ref{sec:topical_diversity}). Next, we explore the accuracy of our assumption on the relevancy of documents (Section \ref{sec:relevance_analysis}). % and investigate how effective our sub-topic generation process is (Section \ref{sec:manual_subtopics}).

%1) \hl{the total of generating our collection. }
%\hl{Zipf distribution nasil - not so important }

%\hl{ap in GTC vs. AP in Combined - olursa iyi olur}

%\hl{p@100 in GTC vs. p@100  in Combined - olursa iyi olur }

%\hl{p@10 in GTC vs. p@10  in Combined  - olmasi lazim}

%\hl{ap with tricky vs. ap without tricky  - olursa iyi olur}

%\hl{tau without random document  - olmasi lazim }

\subsection{Document Structure} \label{sec_ana_doc_str}

\textbf{Table \ref{Tab_stat_1}} presents brief statistics about \testcol, and compares it against the documents in Disks 4-5. Firstly, Disks 4-5 contains around five times more documents than \testcol. This also results in differences in the total number of words (e.g. 41M vs 270M). In addition, the documents and sentences in Disks 4-5 have longer average lengths and higher standard deviations compared to those in \testcol{}.
%however more volatile, which is implied by higher standard deviation in document and sentence lengths.
Therefore, while a document in Disks 4-5  may be extremely long (e.g., document `FR941202-2-00139' has 1,027,658 words), it may also contain only two words (e.g., document `LA011089-0070'). 
%\todo{bunun hangi doküman olduğunu belirtelim. Yani doküman id'si kaç?}. 
Overall, \testcol{} consists of shorter documents and sentences, but the document sizes are more stable compared to human-authored documents. % lengths of these are more steady because all documents are generated with the same process. On the other hand, Disk4-5 documents are less consistent thanks to documents belonging to varying sources.

\begin{comment}
\begin{table}[!htb]
\small
\centering
  \begin{tabular}{ |p{4.8cm} | r | r |}
    \hline
    & \textbf{ChatGPT23} & Disks 4-5 \\ \hline
    The Total Number of Documents & 96,196 & 528,155\\ \hline
 %     The number of topics & 300 & 50 \\ \hline
 %     The number of relevant documents  &  &\\ \hline
 %     The average number of relevant documents per topic  & 63.2 & \\ \hline
    The Total Number of Words  & 41,295,475 & 270,715,891\\ \hline
     The Average Number of Words Per Document  & 429.3 & 516.8 \\ \hline
      Standard Deviation in  Document Length in terms of  words & 112.4 & 1,683.2 \\ \hline
     The Average Number of Sentences Per Document  & 24.0 & 21.5 \\ \hline
      Standard Deviation in  Document Length in terms of  sentences & 7.4 & 39.6 \\ \hline
     The Average Number of Words Per Sentence  & 18.2 & 24.0\\ \hline
     The median of number of words per sentence & 18.0 & 20.0\\ \hline
     Standard Deviation in Number of Words in sentences  & 7.8 & 94.3\\ \hline 
     The Longest Document Length \textbf{kelime mi karakter mi}  & 648 & 822,942**\\ \hline
      The Shortest Document Length   & 6 & 1*\\ \hline
    \hline
  \end{tabular}
  \caption{Statistics about the documents we created. *DOCID XYZ} 
  \label{Tab_stat_1}
\end{table}    
\end{comment}

\begin{table}[!htb]
\small
\centering
\renewcommand{\arraystretch}{1.2}
  \begin{tabular}{ p{9cm}  r  r }
    \hline
    & \textbf{\testcol{}} & \textbf{Disks 4-5} \\ \hline
    The total number of documents & 96,196 & 528,155\\ \hline
 %     The number of topics & 300 & 50 \\ \hline
 %     The number of relevant documents  &  &\\ \hline
 %     The average number of relevant documents per topic  & 63.2 & \\ \hline
    The total number of words  & 44,718,147 & 280,033,874\\ \hline
     The average number of words per document  & 464.86 & 534.54 \\ \hline
     Standard deviation in the number of words \newline per document & 121.95 & 1,866.75 \\ \hline
      %Standard deviation in  document length \newline in terms of words & 121.95 & 1,866.75 \\ \hline
     The average number of sentences per document  & 23.98 & 21.5 \\ \hline
      Standard deviation in the number of sentences \newline per document & 7.43 & 39.6 \\ \hline
     The average number of words per sentence  & 19.39 & 24.84\\ \hline
     The median of the number of words per sentence & 20.0 & 21.0\\ \hline
     Standard deviation in the number of words \newline in sentences  & 8.34 & 105.12\\ \hline 
     Maximum word count in a document   & 700 & 1,027,658\\ \hline
      Minimum word count in a document   & 6 & 2\\ \hline
  \end{tabular}
  \caption{Statistics about the documents in \testcol{} and Disks 4-5.} 
  \label{Tab_stat_1}
\end{table}

\subsection{Lexical Diversity} \label{sec:lexical_diversity}

Using a diverse set of words enriches a document's content and usually reflects the author’s writing capability and creativity. Furthermore, lexical diversity  influences retrieval tasks by reducing textual overlap across documents.  
%While an idea can be expressed via a limited set of words, utilizing synonyms and alternative phrases is a way of reducing repetition and redundancy. Accordingly, lexical is a desired property in written language. 
Therefore, we compare the documents we generated with those in Disks 4–5 in terms of lexical diversity. In particular, we employ the following lexical diversity measures from the literature:
%In particular, we calculate the number of unique lemmas and words in surface level. In addition, we employ the following metrics used in the literature.

\begin{itemize}
    \item \textbf{Type-Token Ratio (TTR)}: The ratio of the number of unique words (types) to the total number of words (tokens) in a text. %\todo{type ile token arasındaki ayrıntı nedir?} 
    A higher TTR indicates greater lexical diversity.
    \item \textbf{Maas TTR} \cite{mass1972zusammenhang}: The logarithmic transformation of TTR designed to reduce the sensitivity to text length. A lower Maas value indicates higher lexical diversity.
    \item \textbf{Hypergeometric Distribution-based D (HDD)} \cite{mccarthy2007vocd}.  HDD calculates the probability that a word chosen at random is unique in a sample, using the hypergeometric distribution to estimate this probability across multiple random samples of different sizes. It’s often considered more robust for comparing texts of different lengths.  
    \item \textbf{Measure of Lexical Textual Diversity (MTLD)} \cite{mccarthy2005assessment}.  MTLD measures the point in a text at which the TTR falls below a specified threshold, resetting the count and continuing until the entire text is analyzed. The final value is the mean length of segments where TTR drops below the threshold. MTLD is also designed to be independent of text length, making it a robust measure for comparing texts of different lengths. A higher MTLD score indicates greater lexical diversity. 
\end{itemize}

 \textbf{Table \ref{tab:lexical_diversity}} presents the results.  We observe that human-authored documents demonstrate greater lexical diversity across all evaluated metrics. This difference is likely attributable to the fact that the Disks 4-5 collection contains documents authored by different writers and sourced from diverse news outlets. In contrast, our proof-of-concept study relies on a single LLM to generate all documents. Lexical diversity in our collection could potentially be enhanced by incorporating multiple LLMs during document generation. Moreover, employing advanced prompt engineering techniques \cite{chen2024genqa} offers another avenue for increasing diversity.  We leave this aspect of our work as future work. 
 
 %On the other hand, all documents in our synthetic collection are generated by only one source, ChatGPT. 
 %For the same reason, the synthetic collection promises consistency in lexical diversity, which can be ensured by lower standard deviation values in the table. \todo{Kendime not: bu paragrafı kısaltabiliriz. bir de hepsinde consistently aynı şeyi söylüyorsa bunu emphasize etsek mi bilemedim. }

\begin{table}[!htb]
\small
\centering
\renewcommand{\arraystretch}{1.2}
  \begin{tabular}{ p{7cm} c c M{0.4cm} c c}
    \hline
    \textbf{Lexical Diversity Metric} & \multicolumn{2}{c}{\bf \testcol} & & \multicolumn{2}{c}{\bf Disks 4-5}  \\ \cline{2-3} \cline{5-6}
     & mean & std & & mean & std \\ \hline
      TTR & 0.39 & 0.12 & & 0.53 & 0.17 \\
      Mass TTR  & 0.06 & 0.01 & & 0.05 & 0.02\\
      HDD  & 0.74 & 0.16 & & 0.77 & 0.17 \\
      MTLD & 60.03 & 11.01 & & 77.51 & 32.17 \\
      Number of unique words per document & 203.6 & 50.9 & & 238.6 & 222.7 \\
      Number of lemmatized unique words \newline per document & 199.3 & 49.7 & & 233.7 & 216.8\\ \hline
  \end{tabular}
  \caption{Lexical diversity of documents in \testcol{}  and Disks 4-5.} 
  \label{tab:lexical_diversity}
\end{table}

%We also calculate Zipf distributions of both collections. The results are shown in \textbf{Figure \ref{fig:zipf_dists}}. We observe that  \testcol{} generally has a similar distribution with Disk 4-5 collection but it has a  smoother distribution than Disk 4-5.

%\begin{figure*}[!htb]
%\centering
%\input{figures/zipf}
%\caption{Zipf distributions of collections}
%\label{fig:zipf_dists}
%\end{figure*}

\subsection{Readability} \label{sec:readability}

Readability is a measure of how easily a document is understood by readers. Various factors influence readability, including the technicality of vocabulary and the complexity of sentence structures. For example, documents containing more technical terms demand higher domain expertise or education. 

To assess the readability of our generated documents, we employ established readability metrics, including Kincaid \cite{kincaid1975derivation}, Flesh Reading Ease (FRE) \cite{flesch1949art} and ARI \cite{senter1967automated}, and compare our collection \testcol{} with Disk 4–5. \textbf{Table \ref{tab:readability}} shows the results. Each metric also provides a corresponding scale that indicates the education level required to comprehend the text. We observe that documents we generated require a higher level of education than those in Disk 4–5 in all evaluation metrics, suggesting that they contain more complex sentences and more technical terms. Similar to  our findings on lexical diversity, the standard deviation in readability scores is lower for our generated documents than for Disks 4–5 documents.

%Readability is a measure of how easily a document is understood by readers. Various factors influence readability, including the technicality of vocabulary and the complexity of sentence structures. For example, as more technical terms are used in a document, it requires more expertise and education in the corresponding field to understand it. To analyze the readability of our generated documents, we employ popular readability metrics from the literature, including Kincaid \cite{kincaid1975derivation}, Flesh Reading Ease (FRE)\cite{flesch1949art} and ARI \cite{senter1967automated} metrics, and compare the documents in \testcol{} with those in Disk 4-5. To interpret the results, the metrics also provide a scale, which relates readability scores to the education level needed to understand the respective document.
%\textbf{Table \ref{tab:readability}} presents the results.  We observe that documents of our collection require a higher education level than the documents of Disk 4-5, suggesting that the generated documents have  more sophisticated sentences and contains more technical terms. Similar to our results for lexical diversity, the generated documents have lower standard deviation score in terms of readability. 

\begin{table}[!htb]
\small
\centering
\renewcommand{\arraystretch}{1.2}
  \begin{tabular}{ l M{0.4cm} ccc M{0.4cm} ccc}
    \hline
    Metric & & \multicolumn{3}{c}{\bf \testcol{}} & & \multicolumn{3}{c}{\bf Disks 4-5}  \\ \cline{3-5} \cline{7-9}
     & & mean & std & level & & mean & std & level \\ \hline
      Kincaid & & 10.2 & 1.58 & 11th grade & & 5.05 & 3.07 & 6th grade \\ 
      FRE  & & 44.61 & 11.98 & college & & 74.31 & 19.79 & 7th grade\\ 
      ARI  & & 11.16 & 1.5 & 10th grade & & 7.25 & 1.94 & 6th grade \\ \hline
  \end{tabular}
  \caption{Readability scores of documents in \testcol{}  and Disks 4-5. Kincaid and ARI scores correlate with the number of years of education someone needs to understand the text, i.e.,  texts with higher scores are more complex. On the other hand, for FRE, higher scores indicate easier reading material.} 
  \label{tab:readability}
\end{table}

\begin{comment}
\begin{table}[!htb]
\small
\centering
  \begin{tabular}{ |l|c|c|c|c|c|c|}
    \hline
    Metric & \multicolumn{3}{c|}{\bf ChatGPT23} &  \multicolumn{3}{c|}{\bf Disc 4-5}  \\ \hline
     & mean & std & level & mean & std & level \\ \hline
      Kincaid & 10.2 & 1.58 & college grad. & 5.05 & 9.41 & professional \\ \hline
      FRE  & 44.61 & 11.98 & college & 74.31 & 391.64 & 7th grade\\ \hline
      ARI  & 11.16 & 1.5 & 10th grade & 7.25 & 3.75 & 6th grade \\ \hline
  \end{tabular}
  \caption{Readability scores of documents in \testcol{}  and Disc 4-5.} 
  \label{tab:readability}
\end{table}    
\end{comment}

\subsection{Topical Diversity} \label{sec:topical_diversity}

A well-designed collection is expected to cover a wide range of topics, thereby improving its representation of real-world diversity. In this section, we compare our collection \testcol{} with Disks 4–5 in terms of topical diversity. We use BERTopic \cite{grootendorst2022bertopic} to identify topics. BERTopic  clusters documents based on their content where each cluster corresponds to one topic.

%A well-designed collection is expected to cover a wide range of topics, thereby enhancing its representation of real-world diversity. Thus, in this section, we compare our collection and Disks 4-5 in terms of topical diversity analysis. We use BERTopic \cite{grootendorst2022bertopic} to identify topics. Topic extraction relies on clustering documents based on their contents. Documents within the same cluster are considered to share a common topic. Thus, the number of clusters corresponds to the amount of topics in the collection.

\begin{figure}[!htb]
    \centering
    \begin{subfigure}[b]{0.45\textwidth}
        \centering
        \includegraphics[scale=0.2]{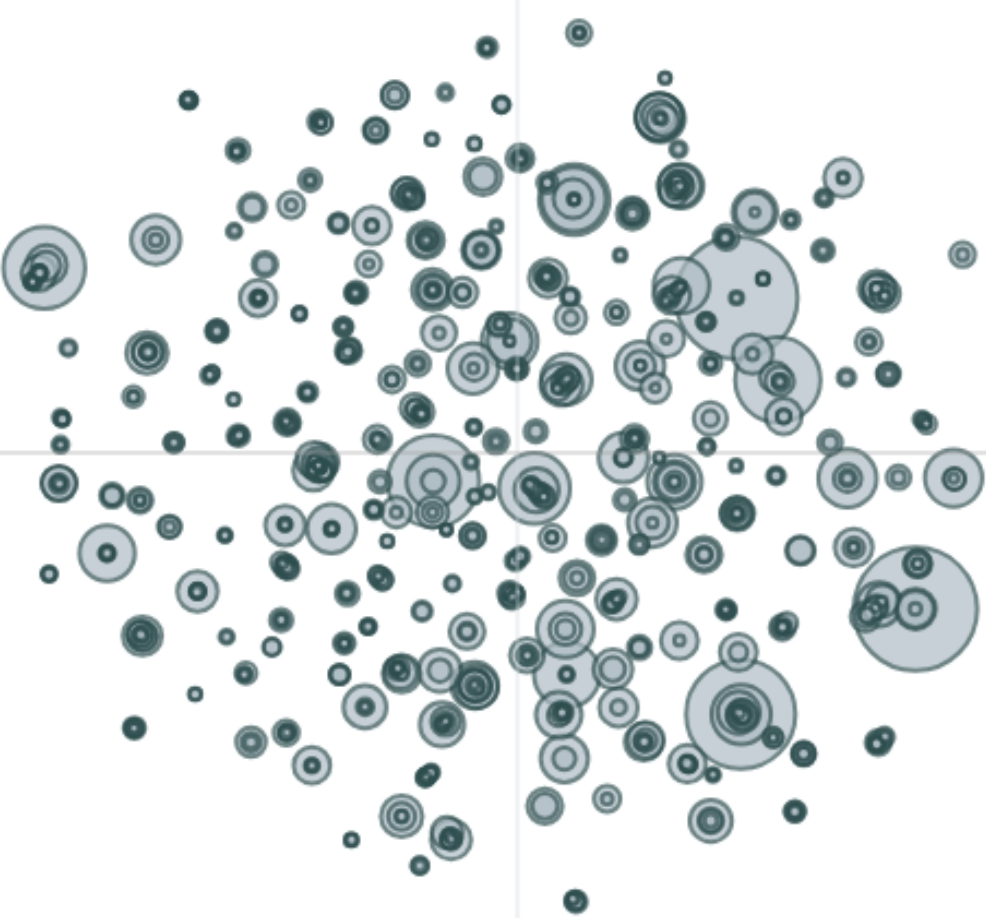}
        \caption{Downsampled Disks 4-5}
        \label{fig:disk45_topical_diversity}
    \end{subfigure}
    \begin{subfigure}[b]{0.45\textwidth}
        \centering
        \includegraphics[scale=0.2]{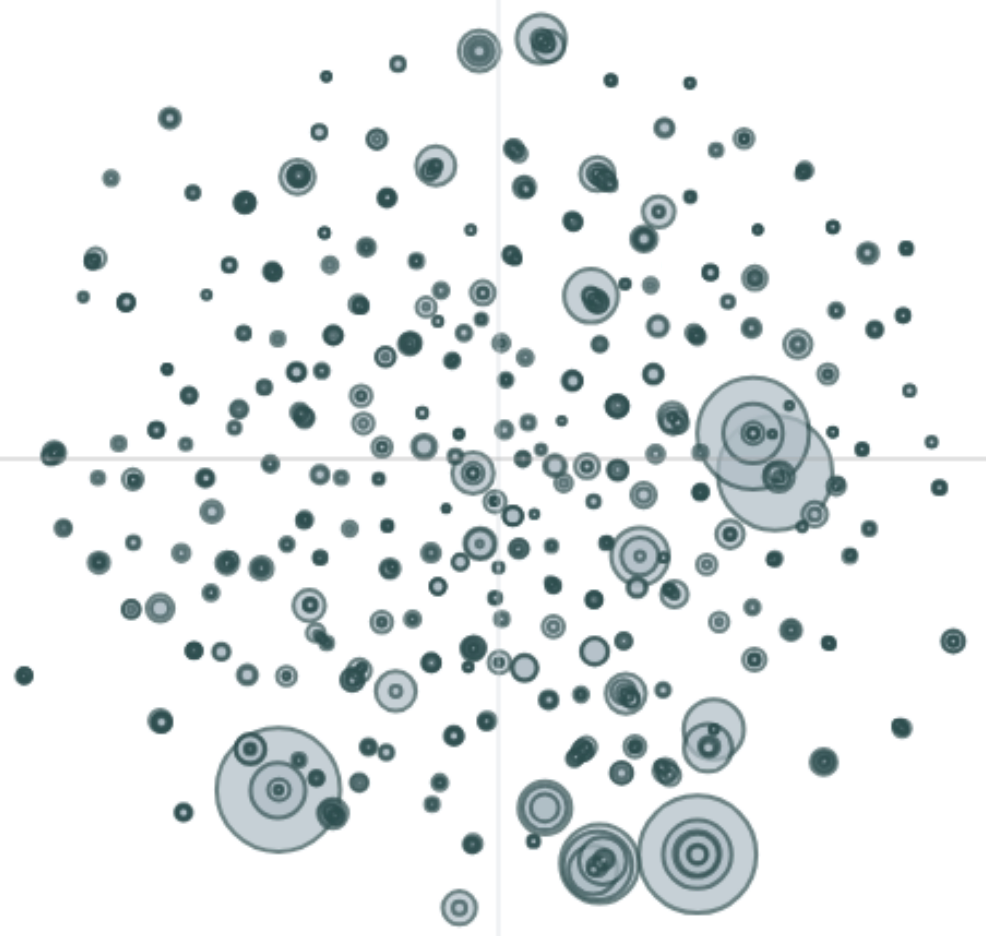}
        \caption{\testcol{}}
        \label{fig:chagpt23_topical_diversity}
    \end{subfigure}
    \caption{Inter-topic Distance Map for \testcol{} and a downsampled version of documents in Disks 4-5. In order to make a fair comparison, we randomly sampled 96,196 documents (i.e., the size of \testcol{}) from Disks 4-5.}
    \label{fig:topical_diversity}
\end{figure}

%While Disks 4-5 collection contains 528,155 documents, we identify 4,674 topics in total. On the other hand, \testcol{} consists of 96,196 documents, covering 945 topics in total. Thus, the number of documents per topic is 112.9 and 101.9 for Disk 4-5 and \testcol{}, respectively. 
While Disks 4–5 has 528,155 documents and 4,674 identified topics, \testcol{} contains 96,196 documents and 945 topics, resulting in 112.9 and 101.9 documents per topic, respectively.
To ensure a fair comparison, we also sampled 96,196 documents from Disks 4-5 and calculated the number of topics within that sample. We repeated this process five times. On average, each sample covered 927 topics. \textbf{Figure \ref{fig:topical_diversity}} illustrates the size of topics (i.e., the number of documents) and their relative distance for \testcol{} and the downsampled Disks 4-5. We observe that the topics in Disks 4-5 are closer to each other than those in \testcol{}, suggesting that \testcol{} covers more different topics than the downsampled Disks 4-5.

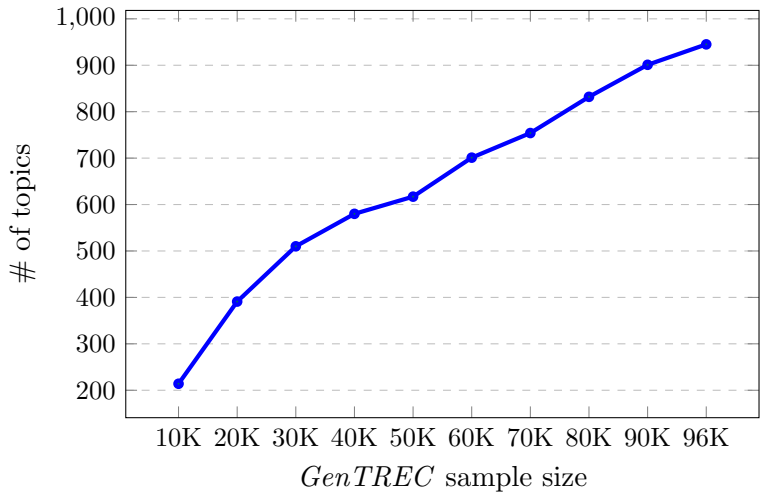
\begin{figure}[!htb]
\centering
\begin{tikzpicture}
\begin{axis}[
    scale= 1,
    width=10cm,
    height=7cm,
    xlabel={\testcol{} sample size},
    ylabel={\# of topics},
    tick label style={font=\small},
    symbolic x coords={10K, 20K, 30K, 40K, 50K, 60K, 70K, 80K, 90K, 96K},
    xtick=data,
    %x tick label style={rotate=45, anchor=east},
    ytick={200,300,...,1000},
    %legend pos=north east,
    legend pos=south east,
    ymajorgrids=true,
    grid style=dashed,
    legend style={nodes={scale=0.8, transform shape}},
    every axis plot/.append style={ultra thick},
    %no markers,
    legend image post style={mark=*},
    cycle list name=exotic,
    mark=*,
    mark options={scale=0.6}
]
\addplot +[blue]
    coordinates {
    (10K,214)(20K,391)(30K,510)(40K,580)(50K,617)(60K,701)(70K,754)(80K,832)(90K,901)(96K,945)
    };
\end{axis}
\end{tikzpicture}
\caption{The number of BERTopic topics wrt varying collection size of \testcol}
\label{fig:topical_diversity_wrt_samples}
\end{figure}

Furthermore, we examined whether we could generate more topics if we had more documents in our collection. Thus, we sampled various numbers of documents from \testcol{} and calculated the number of topics for each sample (\textbf{Figure \ref{fig:topical_diversity_wrt_samples}}). We observe that the number of topics continues to increase as the sample size grows, with no indication of a plateau.
%The results are shown in \textbf{Figure \ref{fig:topical_diversity_wrt_samples}}. We observe that the number of topics continues to grow as additional documents are sampled, with no indication of a plateau. 
This suggests that the number of topics is likely to expand further with more documents.  Overall,  our results suggest that we can create a document collection covering a diverse set of topics using LLMs. %\todo{eğer submission'ı geciktirmeyecekse: aslında burada topic sayısındaki artış eğilimini göstermek iyi olabilir. doküman sayısı 10K,20K,...100K iken topic sayısı nasıl. Eğer burada bir plato oluşmaya başlıyorsa o kadar sayıda topic'e ulaşamayabiliriz. Ama eğer hala artış devam ediyorsa dediğimiz gibi olabilir.}

%While Disk 4-5 collection covers 4,674 topics, ChatGPT23 consists of 945 topics in total. The distribution of topics spaces for regarding collections is illustrated in \textbf{Figure \ref{fig:disk45_topical_diversity}} and \textbf{Figure \ref{fig:chagpt23_topical_diversity}}. 
%The most probable reason behind this is the size of the collections. There exist 528,155 documents and 96,196 documents in Disk4-5 and ChatGPT23 collections, respectively. In this case, the number of documents per topic is 112.9 and 101.9 for Disk4-5 and ChatGPT23 collections. This indicates that our method of collection generation is efficient in that it yields more topics with the same amount of documents. For a fair comparison, we equalized the number of documents in collections, by sampling 96,196 from Disk4-5. We use 5 as the randomization factor. 
%The 8 topics covered the most by the documents for both collections are depicted in \textbf{Figure \ref{fig:largets_topics}}. In the figure, topics are represented and described with keywords. 

%While Disk4-5 contains 500,000 documents, ChatGPT23 contains 85,712 documents
%In this section, Upon extracting topics, we observe how two collections (CHATGPT23 and Disk4-5) differ in topical diversity. As a result, 

\subsection{Accuracy of Relevance Judgments} \label{sec:relevance_analysis}

In this section, we evaluate our methods for creating relevant documents and “tricky” non-relevant documents.
To conduct this evaluation, we randomly sampled 10 topics from TREC-8 and one of the authors manually annotated the generated documents categorized as relevant or tricky non-relevant for each sampled topic. \textbf{Table \ref{tab:relevance_analysis}} presents the relevance rate for documents in the relevant category and the non-relevance rate for the documents in the tricky non-relevant category.

\begin{table}[!htb]
\small
\centering
\renewcommand{\arraystretch}{1.2}
  \begin{tabular}{ c M{5cm} M{5cm} }
    \hline
    Topic ID & Relevance Rate in Relevant-Judged Documents &  Non-Relevance Rate Tricky Non-Relevant Documents \\ \hline
    402 & 0.67 (33/49) & 1.00 (60/60) \\ \hline
    416 & 1.00 (42/42) & 1.00 (66/66) \\ \hline
    417 & 0.97 (61/63) & 1.00 (43/43) \\ \hline
    419 & 0.98 (47/48) & 1.00 (42/42) \\ \hline
    429 & 0.86 (25/29) & 0.91 (60/66) \\ \hline
    430 & 0.00 (0/48) & 0.87 (52/60) \\ \hline
    434 & 1.00 (86/86) & 0.97 (64/66) \\ \hline
    437 & 1.00 (68/68) & 0.64 (27/42) \\ \hline
    438 & 0.80 (35/44) & 1.00 (60/60) \\ \hline
    441 & 0.85 (67/79) & 0.98 (59/60) \\ \hline \hline
    \textbf{Average} & \textbf{0.83} & \textbf{0.94} \\ \hline
  \end{tabular}
  \caption{Accuracy of our method for relevant and tricky non-relevant documents for 10 topics. The numbers within parentheses show the number of correctly ``judged'' documents and the total number of documents for each category.} 
  \label{tab:relevance_analysis}
\end{table}

Our findings indicate an average accuracy of 83\% in generating relevant documents and 94\% in generating tricky non-relevant documents. However, performance in generating relevant documents varies substantially across topics. Specifically, for five topics (416, 417, 419, 434, 437), we achieve over 97\% relevance accuracy. For the remaining topics, our method demonstrates acceptable performance, except for Topic 430. Surprisingly, none of the documents intended to be relevant for Topic 430 (which focuses on incidents of Africanized bee attacks on humans) were actually relevant. Upon closer examination, we found that although the generated documents addressed certain aspects of killer bees (e.g., origin, physical characteristics, and ecological impact), they did not reference specific historical attacks, resulting in a 0\% relevance rate.
Thus,  we further investigated whether our method systematically underperforms on topics requiring references to specific past incidents, as in the case of Topic 430. We identified seven additional TREC-8 topics of this nature\footnote{The ID of these topics are 408, 422, 424, 425, 429, 442, and 448.} and manually assessed the relevance of the generated documents. For these topics, the average relevance rate was 0.37, indicating that our method struggles with such topics. Addressing this limitation requires further exploration, which we leave as future work.

%With Topic 402, the first 35 documents are mostly relevant. However, the relevance rate decreases drastically after the 35th document, causing a comparably low relevance rate for the regarding topic.

Topic 402 also yielded a relatively low relevance rate (0.67). Regarding this topic, we observed that the first 35 generated documents were mostly relevant, but the relevance rate decreased significantly beyond this point, lowering the overall relevance rate for this topic. This implies that the optimal number of relevant documents (or subtopics) may vary by topic, indicating a need for future work on how to detect the most suitable number of subtopics to generate.

%We observed a higher success rate, averaging 0.94, in generating tricky-non-relevant documents. Except Topic 430 and 437, we achieved an accuracy in generating tricky-non relevant documents higher than 91\%. Interestingly, for 'Topic 430', the non-relevance rate is not 1.0, indicating that some documents intended to be irrelevant are actually relevant. Although our approach did not yield any relevant documents when targeting relevance, it did produce some relevant documents when targeting non-relevance.
Our method demonstrates a higher performance in generating tricky non-relevant documents, with an average success rate of 94\%. We again observe that the success of our method varies across topics. For instance, while we achieve 100\% non-relevance rate in five topics, Topic 437 exhibits a noticeably lower accuracy (0.64). Therefore, determining how to identify topics best suited for generating tricky non-relevant documents remains an open question for further exploration.

\section{System Evaluation with \testcol{}} \label{sec:evaluation}

In this section, we investigate whether we can use \testcol{} to reliably evaluate IR systems. We first explain our experimental setup (Section \ref{sec_exp_setup}). Next, we present our results, seeking answers for our research questions (Section \ref{sec_exp_results}).
%explain the experimental setup (Section X). Subsequently, we seek answers for the research questions listed in Section \ref{sec_exp_results}. 

\subsection{Experimental setup} \label{sec_exp_setup}

%\subsubsection{Text Generation}

%\subsubsection{Evaluation Metrics}
%Tau'yu kullandık 
%vektimiz olursa tau ap kullanabiliriz. kullanırsak tau ap burada anlatılmalı

%Recall, P@10, map, ndcg kullanıldı. 
%Trec eval kullanıldu

%In order to rank systems, 5 different metrics have been used such as MAP, Precision @ 10, NDCG, recall @ 1000, RPREC.

\subsubsection{Systems Used In Experiments}

To assess the evaluation reliability of \testcol{},  we implemented various retrieval systems using  Pyserini library \cite{Lin_etal_SIGIR2021_Pyserini}.  Specifically, we developed a lexical retrieval model, two dense retrieval models, and a re-ranker model, which were combined in various configurations to obtain additional systems. For lexical retrieval, we use BM25, a classic bag-of-words approach that incorporates the Porter stemmer. The two dense-retrieval models are \textit{aggretriever-distilbert} \cite{lin2023aggretriever} and \textit{tct\_colbert-v2-hnp-msmarco} \cite{lin2021batch}. Additionally, we use \textit{monot5-base-msmarco} \cite{nogueira2020document} as our re-ranker model. 
%In order to augment the number of systems, we use various combinations of retrieval models (e.g., multistage retrieval). For instance, BM25 and aggretriever-distilbert are two separate retrieval models. As an additional retriever, we create a pipeline where BM25 retrieves 10,000 documents, which are then passed to aggretriever-distilbert, where it selects 100 documents from the initial set. In this way, we obtain an additional system which is different from BM25 and aggretriever-distilbert. \textbf{Table \ref{tab_ir_systems}} shows ten IR systems used in this study.
We use Pyserini's ``hybrid search" module to combine different retrieval models. For example, BM25 and \textit{aggretriever-distilbert} serve as two independent retrieval models. By merging their scores, we create a third hybrid retriever, yielding an additional system distinct from BM25 and \textit{aggretriever-distilbert}. 
Furthermore, we enhance system diversity by incorporating  MonoT5 re-ranker. This approach generates alternative systems by re-ranking the documents retrieved by other models. In total, we implemented 10 different systems. \textbf{Table \ref{tab_ir_systems}} shows these IR systems used in this study.

\begin{table*}[!htb]
\scriptsize
\centering
\renewcommand{\arraystretch}{1.2}
  \begin{tabular}{ l p{6.2cm} }
    \hline
    \textbf{Retrieval Model} & \textbf{Description}\\
    \hline
    \hline
    BM25 & The popular lexical retrieval model with Porter Stemmer \\ \hline
    aggretriever-distilbert\cite{lin2023aggretriever} & A single-vector dense retrieval model where all contextualized token embeddings provided to BERT as input.  \\ \hline
    tct\_colbert-v2-hnp-msmarc \cite{lin2021batch} & A dense retrieval model applying knowledge distillation using the ColBERT late-interaction  model.   \\ \hline
    bm25 + aggretriever-distilbert & The hybrid model using both BM25 and aggretriever-distilbert model  \\ \hline
    bm25 + tct\_colbert-v2-hnp-msmarco &  The hybrid model using both BM25 and tct\_colbert-v2-hnp-msmarco \\ \hline
    bm25 + MonoT5 & The results of BM25 are reranked by MonoT5 \\ \hline
    aggretriever-distilbert + MonoT5 &  The results of aggretriever-distilbert are reranked by MonoT5 \\ \hline
    tct\_colbert-v2-hnp-msmarco + MonoT5 &  The results of aggretriever-distilbert are reranked by MonoT5 \\ \hline
    bm25 + aggretriever-distilbert + MonoT5 & The results of the hybrid model (bm25 + aggretriever-distilbert) are reranked by  MonoT5\\ \hline
    bm25 + tct\_colbert-v2-hnp-msmarco + MonoT5 &  The results of the hybrid model (bm25 + tct\_colbert-v2-hnp-msmarco) are reranked by  MonoT5 \\ \hline
  \end{tabular}
  \caption{ The IR systems we use in our evaluations.% Lexical model corresponds to the classical bag-of-words approach. With dense retrieval models, dense representations (e.g. embeddings) of documents and queries are extracted, and documents are retrieved based on their embedding similarity to the query embedding. Separate encoders are employed for documents and queries. Rerankers works on a limited document list and ranks documents based on which document is more relevant to the query.
  } 
  \label{tab_ir_systems}
\end{table*}

\subsubsection{Test Collections}
We evaluate our IR systems on our test collection and also several existing TREC collections for comparative analysis. In particular, we use TREC-5 \cite{trec5}, TREC-6 \cite{trec6}, TREC-7 \cite{trec7}, TREC-8 \cite{trec8-overview}, Robust2004 \cite{robust2004} collections as baselines. Furthermore, since all these TREC collections use the same document collection, we artificially create another test collection, named TREC\textsubscript{All} that contains all TREC topics and the respective relevance judgments. %\todo{bunu da table'a ekleyelim}
We selected these test collections because we generated documents using their topics, allowing us to conduct a fair comparison between \testcol{} and them. \textbf{Table \ref{Tab_test_collections}} presents general statistics for all test collections used in this study.

\begin{table*}[!htb]
\footnotesize
\centering
\renewcommand{\arraystretch}{1.2}
  \begin{tabular}{ p{3cm}  r  r  r  r  r  r r }
    \hline
    & \textbf{\testcol{}} & \textbf{TR5}  & \textbf{TR6}  & \textbf{TR7}  & \textbf{TR8}  &  \textbf{R04} &  \textbf{TREC\textsubscript{All}}  \\ \hline
      The number of \newline topics & 300 & 50 & 50 & 50 &50  & 249 & 299 \\ \hline
      The number of \newline relevant documents  & 18,964 & 5,524 & 4,611 & 4,674 & 4,728 &   16,381 &  22,226\\ \hline
      The number of unique relevant documents  & 18,964 & 5,306 & 4,482 & 4,555 & 4,628 &  15,020 &  20,360\\ \hline
      Relevant document ratio  & 0.197 & 0.010 & 0.009 & 0.009 & 0.009 & 0.031 &  0.042\\ \hline
%      Unique Relevant document ratio  & 0.197 & 0.010 & 0.008 & 0.009 & 0.009 & 0.028 & 0.039 \\ \hline
      The average \newline number of relevant documents per topic  & 63.2 & 110.5 & 92.2 & 93.5 & 94.6 &   66.1 &  74.6\\ \hline
      The std of \newline relevant documents per topic  & 16.1 & 139.1 & 103.1 & 85.2 &  80.0 &   75.6 &  92.1\\ \hline
  %    The min of relevant documents per topic  & \hl{2} & 1 & 3 & 7 & 6 &  0 & 0\\ \hline
      The max of \newline relevant documents per topic  & 100 & 594 & 474 & 361 & 347 &   448 &  594\\ \hline
  \end{tabular}
  \caption{Statistics about \testcol{} and the test collections we use in our experiments. } 
  \label{Tab_test_collections}
\end{table*}%\todo{biz de sadece 2 tane relevant document olan topic nasıl oldu?}

\subsubsection{Evaluation Metrics}

\begin{comment}

\begin{table}[!ht]
    \small
    \centering
    \begin{tabular}{|l|l|l|l|l|}
    \hline
        ndcg\_tau & trec-5 & trec-6 & trec-7 & trec-8 \\ \hline
        trec-5 & 1.000 & 0.687 & 0.597 & 0.735 \\ \hline
        trec-6 & 0.687 & 1.000 & 0.739 & 0.741 \\ \hline
        trec-7 & 0.597 & 0.739 & 1.000 & 0.621 \\ \hline
        trec-8 & 0.735 & 0.741 & 0.621 & 1.000 \\ \hline
    \end{tabular}
    \caption{Tau correlation between system run results using NDCG on TREC collection} 
\end{table}

\begin{table}[!ht]
    \small
    \centering
    \begin{tabular}{|l|l|l|l|l|}
    \hline
        rprec\_tau & trec-5 & trec-6 & trec-7 & trec-8 \\ \hline
        trec-5 & 1.000 & 0.670 & 0.566 & 0.753 \\ \hline
        trec-6 & 0.670 & 1.000 & 0.713 & 0.713 \\ \hline
        trec-7 & 0.566 & 0.713 & 1.000 & 0.628 \\ \hline
        trec-8 & 0.753 & 0.713 & 0.628 & 1.000 \\ \hline
    \end{tabular}
    \caption{Tau correlation between system run results using RPREC on TREC collection} 
\end{table}

\end{comment}

In our experiments, we use five metrics to quantify the performance of IR systems, including Precision@10 (P@10), Precision@100 (P@100), mean average precision (MAP), and R-Precision. We compare rankings of IR systems in \testcol{} against rankings in TREC collections to assess the evaluation reliability. We use Kendall's $\tau$ for ranking comparisons. 
%while evaluating the 10 IR systems we developed (Table \ref{tab_ir_systems}). One of them is Precision which is a set-based metric. We use Precision with two different cut-off thresholds (e.g. 10 and 100), which yields two variants of the metric. The rest 3 of them are rank-based metrics, which are MAP, nDCG \cite{jarvelin2017ir} and R-Precision.

%First and foremost, systems have been created using TREC test collection that occurred in trec-5, trec-6, trec-7, and trec-8. Even though each TREC[5-8] has the same test collection, their topics differ. Therefore, our intention is to measure these 30 systems using the above 5 metrics in TREC test collection with 4 different sets of topics.

\begin{comment}
    
\begin{figure}
  \centering
    \includegraphics[width=0.5\textwidth]{figures/trec8-run-results.pdf}
     \caption{System performances in TREC test collection using TREC-8's topics}
      \label{fig_manuel}
\end{figure}
\end{comment}

%\subsubsection{baselines}
%- Naive approach: topic'i al, metin oluşturt. Sonra re-generate de sürekli. 
%- Real test collections

\subsection{Experimental Results} \label{sec_exp_results}

\subsubsection{Rank Correlation across Test Collections} \label{sec:chatgpt_vs_disk45}

\testcol{} differs from traditional test collections in several aspects, such as its construction methodology, the volume of documents, the number of topics, and the distribution of relevant documents per topic. As these factors can significantly impact IR system evaluations, we start by analyzing the variations in IR system rankings across different test collections, including \testcol{} and the five selected TREC collections. Specifically, we evaluate the IR systems we developed on \testcol{} and each of the five TREC collections independently, using four evaluation metrics to generate system rankings. Subsequently, we compare the rankings derived from \testcol{} with those from the TREC collections. We also analyze how system rankings change across the TREC collections. This allows us to observe the impact of topic set changes on system rankings, as the only difference across the selected TREC collections are  the topic sets. \textbf{Table \ref{tab_tau_map_rprec_prec}} presents the results for MAP, RPREC, P@100, and P@10, separately.

\begin{table}[H]
    \scriptsize
    \centering
    \renewcommand{\arraystretch}{1.2}
         \begin{tabular}{ c c c c c c c c }
        \hline
        & \multicolumn{7}{c}{\textbf{MAP}} \\ \hline
         & \testcol{} & TREC5 & TREC6 & TREC7 & TREC8 & R04 & TREC\textsubscript{All} \\
        \testcol{} & - & 0.51 & 0.69 & 0.78 & 0.82 & 0.73 & 0.78 \\
        TREC5 & 0.51 & - & 0.38 & 0.56 & 0.69 & 0.51 & 0.56 \\
        TREC6 & 0.69 & 0.38 & - & 0.73 & 0.6 & 0.87 & 0.82 \\
        TREC7 & 0.78 & 0.56 & 0.73 & - & 0.78 & 0.87 & \textbf{0.91} \\
        TREC8 & 0.82 & 0.69 & 0.6 & 0.78 & - & 0.73 & 0.78 \\
        R04 & 0.73 & 0.51 & 0.87 & 0.87 & 0.73 & - & \textbf{0.96} \\
        TREC\textsubscript{All} & 0.78 & 0.56 & 0.82 & \textbf{0.91} & 0.78 & \textbf{0.96} & -\\ \hline
        \textbf{Average} & 0.72 & 0.53 & 0.68 & 0.77 & 0.73 & 0.78 & 0.8 \\ \hline
        \end{tabular}

        \vspace{0.5cm}
        
        \begin{tabular}{ c c c c c c c c }
        \hline
        & \multicolumn{7}{c}{\textbf{RPrec}}  \\  \hline
         & \testcol{} & TREC5 & TREC6 & TREC7 & TREC8 & R04 & TREC\textsubscript{All} \\
        \testcol{} & - & 0.75 & 0.85 & 0.85 & 0.75 & 0.85 & 0.85 \\
        TREC5 & 0.75 & - & 0.67 & 0.76 & 0.84 & 0.76 & 0.76 \\
        TREC6 & 0.85 & 0.67 & - & 0.82 & 0.81 & \textbf{0.91} & \textbf{0.91} \\
        TREC7 & 0.85 & 0.76 & 0.82 & - & 0.81 & \textbf{0.91} & \textbf{0.91} \\
        TREC8 & 0.75 & 0.84 & 0.81 & 0.81 & - & \textbf{0.9} & \textbf{0.9} \\
        R04 & 0.85 & 0.76 & \textbf{0.91} & \textbf{0.91} & \textbf{0.9} & - & \textbf{1.0} \\
        TREC\textsubscript{All} & 0.85 & 0.76 & \textbf{0.91} & \textbf{0.91} & \textbf{0.9} & \textbf{1.0} & - \\ \hline
        \textbf{Average} & 0.82 & 0.76 & 0.83 & 0.85 & 0.83 & 0.89 & 0.89 \\ \hline
        \end{tabular}

        \vspace{0.5cm}

        \begin{tabular}{ c c c c c c c c }
        \hline
        & \multicolumn{7}{c}{\textbf{P@100}}  \\  \hline
         & \testcol{} & TREC5 & TREC6 & TREC7 & TREC8 & R04 & TREC\textsubscript{All} \\
        \testcol{} & - & 0.77 & \textbf{1.0} & 0.86 & 0.77 & \textbf{0.95} & \textbf{0.95} \\
        TREC5 & 0.77 & - & 0.77 & 0.81 & \textbf{0.91} & 0.81 & 0.81 \\
        TREC6 & \textbf{1.0} & 0.77 & - & 0.86 & 0.77 & \textbf{0.95} & \textbf{0.95} \\
        TREC7 & 0.86 & 0.81 & 0.86 & - & \textbf{0.91} & \textbf{0.91} & \textbf{0.91} \\
        TREC8 & 0.77 & \textbf{0.91} & 0.77 & \textbf{0.91} & - & 0.81 & 0.81 \\
        R04 & \textbf{0.95} & 0.81 & \textbf{0.95} & \textbf{0.91} & 0.81 & - & \textbf{1.0} \\
        TREC\textsubscript{All} & \textbf{0.95} & 0.81 & \textbf{0.95} & \textbf{0.91} & 0.81 & \textbf{1.0} & - \\ \hline
        \textbf{Average} & 0.88 & 0.81 & 0.88 & 0.88 & 0.83 & \textbf{0.91} & \textbf{0.91} \\ \hline
        \end{tabular}

        \vspace{0.5cm}
        
        \begin{tabular}{ c c c c c c c c }
        \hline
        & \multicolumn{7}{c}{\textbf{P@10}}  \\ \hline
         & \testcol{} & TREC5 & TREC6 & TREC7 & TREC8 & R04 & TREC\textsubscript{All} \\
        \testcol{} & - & -0.1 & 0.23 & 0.32 & 0.23 & 0.3 & 0.23 \\
        TREC5 & -0.1 & - & 0.71 & 0.61 & 0.66 & 0.63 & 0.71 \\
        TREC6 & 0.23 & 0.71 & - & 0.75 & 0.8 & 0.85 & 0.89 \\
        TREC7 & 0.32 & 0.61 & 0.75 & - & 0.89 & 0.85 & 0.84 \\
        TREC8 & 0.23 & 0.66 & 0.8 & 0.89 & - & \textbf{0.94} & \textbf{0.93} \\
        R04 & 0.3 & 0.63 & 0.85 & 0.85 & \textbf{0.94} & - & \textbf{0.94} \\
        TREC\textsubscript{All} &  0.23 & 0.71 & 0.89 & 0.84 & \textbf{0.93} & \textbf{0.94} & - \\ \hline
        \textbf{Average} & 0.2 & 0.54 & 0.7 & 0.71 & 0.74 & 0.75 & 0.76\\ \hline
        \end{tabular}

        \caption{Kendall's $\tau$ correlation across test collections using MAP, RPrec, P@100, and P@10. The correlation scores higher than 0.90 are written in \textbf{bold}.}
        \label{tab_tau_map_rprec_prec}
\end{table}

Firstly, we observe that rank correlations vary across both test collections and metrics. Notably, TREC-5 produces the lowest correlations on all metrics except P@10. Meanwhile, \testcol{} achieves higher Kendall’s $\tau$ scores than TREC-5 in three of the four metrics, showing comparable results to TREC-6, TREC-7, and TREC-8. Only R04 and TREC\textsubscript{All} achieve consistently higher $\tau$ scores than \testcol{}, likely because they are constructed by combining topics from other TREC collections and therefore exhibit higher similarity to those collections.

%Firstly, we observe that ranking correlations vary across collections and metrics. While TREC-5 produces the lowest correlations on all metrics, except P@10. It is noteworthy that \testcol{} achieves a higher Kendall's $\tau$ score than TREC-5 in three metrics, highly compatible results with TREC-6, TREC-7, and TREC-8. %Although TREC collections share the same document collection (e.g. Disk4-5), while we use completely a different document collection. 
%Only R04 and TREC\textsubscript{All} achieve consistently higher $\tau$ score than \testcol{}. This is because these test collections have high similarity with other test collections as they are just constructed by just combining topics existing in other collections. 

Next, focusing on \testcol{}, P@100 yields the highest correlation on average. In particular, against the TREC\textsubscript{All} collection, Kendall’s $\tau$ score reaches 0.95 when using P@100. However, P@10 diverges from the other metrics, producing the lowest correlation scores. This might be because \testcol{}’s average relevant document ratio is much higher than that of other TREC collections (see Table \ref{Tab_stat_1}), allowing systems to retrieve many relevant documents in the top 10 and thus limiting P@10’s ability to differentiate among systems. We observe a similar pattern in the TREC collections, where P@100 again yields the most similar rankings and P@10 the least.

%Secondly, among results for \testcol{}, P@100 yields the highest correlation on average. Specifically, in TREC\textsubscript{All} collection, we achieve 0.95 Kendall's $\tau$ score based on P@100.  
%However, P@10 exhibits an obvious diversion from other metrics, producing the lowest correlation scores. This might be because the average relevant document ratio in \testcol{} is much higher than the other TREC collections (See Table \ref{Tab_stat_1}), enabling systems to retrieve many relevant documents in the top 10 documents, and thereby, not being able to differentiate performance of systems well enough. Furthermore, we observe that TREC collections produce the most similar rankings with P@100 and the lowest correlations with P@10, similar to our results with \testcol{}.

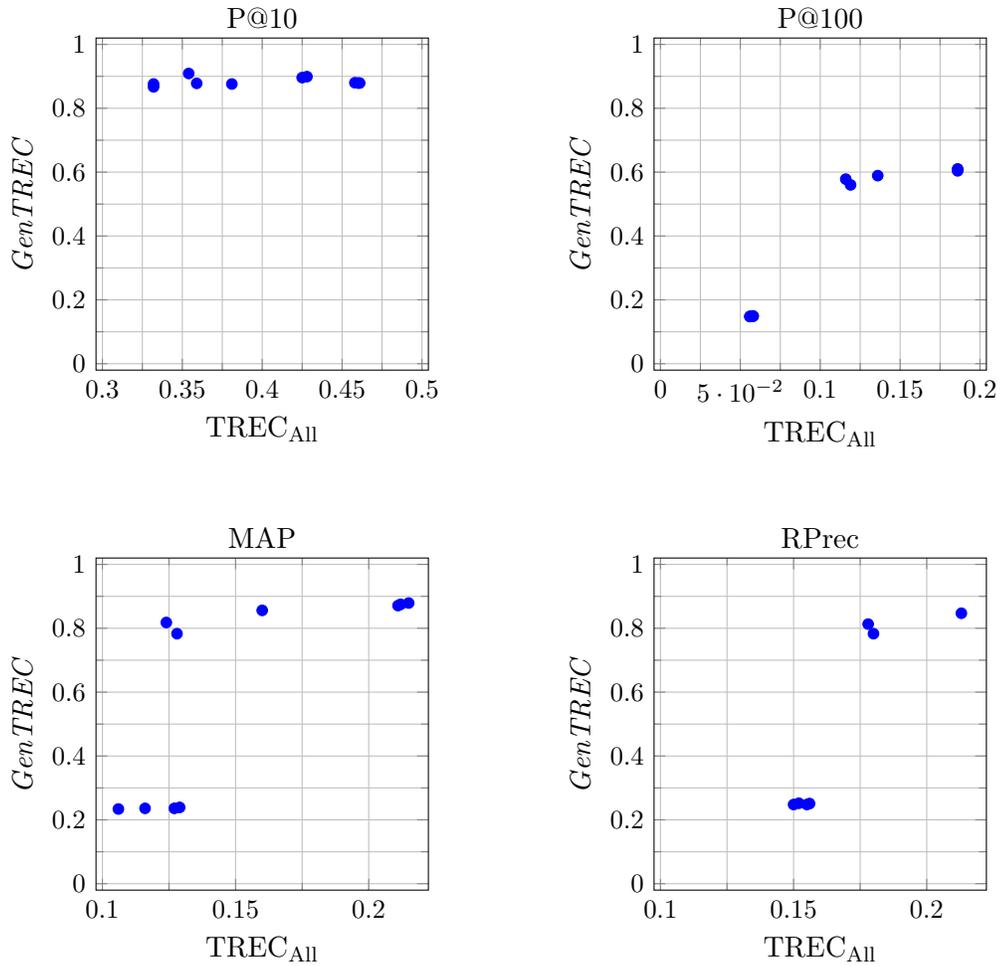
\begin{figure*}[!htb]
\centering
\pgfplotsset{% global config
%  compat=newest,
   grid=both,
  %try min ticks=3,
  minor tick num=1,
  enlargelimits=0.02,
   every tick label/.append style={font=\small},
  group style={
    columns=4,
    xlabels at=edge bottom,
    ylabels at=edge left},
  every axis legend/.append style={
    legend cell align=left,
%    legend columns=7
  }
}

\begin{tikzpicture}
\begin{groupplot}[group style={group size= 2 by 2, horizontal sep=3cm, vertical sep=2.5cm},
       % grid style          = {line width=.1pt,draw=gray!10},
       % major grid  style={line width=.2pt,draw=gray!50},
        height              = 6cm, 
        width=6cm,
        %xlabel              = $\alpha$ ,
        %legend entries={AP vs. $AP_{Rareness}$, P@100 vs. $P@100_{Rareness}$ } ,
        legend style={
            at={(1.5,4)},
            anchor=north west,
            legend columns=-1,transpose legend,
            nodes={font=\small},
            /tikz/every even column/.append style={column sep=0.1cm}
        },
        grid=both,   
	    yticklabel style    = {/pgf/number format/precision=4}  ,
	    scaled y ticks      = false,
         every axis title shift=0,
      legend to name=grouplegend,
        ]

\nextgroupplot[title=P@10,
 ylabel style={align=center}, 
ylabel={\testcol},ylabel shift = 0 pt, xlabel={TREC\textsubscript{All}}, ymin=0,ymax=1, xmin=0.3, xmax=0.5]   
 \addplot +[only marks, mark options={scale=1} ] coordinates {(0.332,0.867)(0.332,0.876)(0.359,0.878)(0.425,0.896)(0.461,0.879)(0.354,0.909)(0.381,0.876)(0.458,0.88)(0.428,0.899)(0.46,0.879)};

\nextgroupplot[title=P@100,
 ylabel style={align=center},
 ylabel={\testcol},ylabel shift = 0 pt, xlabel={TREC\textsubscript{All}}, ymin=0,ymax=1, xmin=0, xmax=0.2]   
 \addplot +[only marks, mark options={scale=1} ] coordinates {(0.119,0.56)(0.116,0.578)(0.056,0.148)(0.056,0.148)(0.186,0.604)(0.136,0.589)(0.058,0.149)(0.186,0.61)(0.058,0.149)(0.186,0.607)};

\nextgroupplot[title=MAP,
 ylabel style={align=center}, 
ylabel={\testcol},ylabel shift = 0 pt, xlabel={TREC\textsubscript{All}}, ymin=0,ymax=1, xmin=0.1, xmax=0.22]   
 \addplot +[only marks, mark options={scale=1} ] coordinates {(0.128,0.783)(0.124,0.818)(0.106,0.234)(0.127,0.236)(0.211,0.871)(0.16,0.856)(0.116,0.236)(0.215,0.879)(0.129,0.239)(0.212,0.875)};

\nextgroupplot[title=RPrec,
 ylabel style={align=center}, 
ylabel={\testcol},ylabel shift = 0 pt, xlabel={TREC\textsubscript{All}}, ymin=0,ymax=1, xmin=0.1, xmax=0.22]   
 \addplot +[only marks, mark options={scale=1} ] coordinates {(0.18,0.783)(0.178,0.813)(0.15,0.248)(0.155,0.248)(0.274,0.872)(0.213,0.847)(0.152,0.252)(0.278,0.878)(0.156,0.251)(0.276,0.875)};
 
%\nextgroupplot[title=MAP,
% ylabel style={align=center}, 
%ylabel={\testcol},ylabel shift = 0 pt, xlabel={\testcol{} w/o TNR}, ymin=0.2,ymax=1, xmin=0.2, xmax=1]   
% \addplot +[only marks, mark options={scale=1} ] coordinates {(0.893,0.783)(0.92,0.818)(0.249,0.234)(0.249,0.236)(0.959,0.871)(0.932,0.856)(0.251,0.236)(0.968,0.879)(0.252,0.239)(0.964,0.875)};

\end{groupplot}

%\node (a) at (11,1.4) {\rotatebox{270}{$P@100_{Rareness}$}};  

    \ref{grouplegend}
\end{tikzpicture}
\caption{Scores for each system in \testcol{} vs. TREC\textsubscript{All}.  }
\label{fig:system_score_scatters}
\end{figure*}

To further investigate system rankings in \testcol{}, we compare each system’s scores in \testcol{} and TREC\textsubscript{All} across the four metrics. \textbf{Figure \ref{fig:system_score_scatters}} presents the results. Our findings show that all systems achieve higher scores on \testcol{} than on TREC\textsubscript{All}. In addition, the system scores in P@10 lie within a narrow range, with every system exceeding 0.8 on \testcol{}, while their P@10 scores range between 0.3 and 0.5 on TREC\textsubscript{All}. This suggests that \testcol{} contains several ``easy" documents to retrieve, thus weakening P@10’s ability to distinguish performance differences. Nevertheless, the rankings based on other metrics remain generally consistent, with only slight shifts for some systems. Indeed, the top two performing systems are the same in both \testcol{} and TREC\textsubscript{All} under these three metrics.

%In order to further investigate how systems are ranked in \testcol{}, we compare each system's score at \testcol{} and  TREC\textsubscript{All} based on four metrics we use. \textbf{Figure \ref{fig:system_score_scatters}} presents the results.
%We observe that all systems achieve higher scores in \testcol{} compared to TREC\textsubscript{All}. In addition, all systems have more than 0.8 P@10 score and their scores are similar while systems achieve lower P@10 scores in TREC\textsubscript{All} and the scores are distributed within the range of 0.3 and 0.5. This suggests that there were several easy documents to be retrieved in \testcol{}, reducing its capability to distinguish their performance based on P@10. However, based on other rankings, we do observe that rankings are generally similar with small rank differences for some systems. Furthermore, we observe that the best performing two systems are the same in both collections based on these three metrics. 

Overall, \testcol{} produces rankings similar to those of TREC\textsubscript{All} in P@100, MAP, and RPrec metrics, successfully 
 identifying the same best-performing systems as observed in TREC\textsubscript{All}. Minor ranking differences are expected, given that \testcol{} consists of different topics and documents from those in TREC collections. However, it leads to substantially different evaluations for P@10, which requires further investigation and possibly different approaches to generate relevant documents. We leave this direction of research as future work.

\subsubsection{Impact of Tricky Non-relevant Documents} \label{sec:impact_tnr}

\begin{figure}[!htb]
\centering
\begin{tikzpicture}
\begin{axis}[
    scale= 1,
    width=8cm,
    height=6cm,
    xlabel={TREC Collections},
    ylabel={Kendall's tau},
    ymin=-0.2, ymax=1.05,
    symbolic x coords={TREC5, TREC6, TREC7, TREC8, R04, TREC\textsubscript{All}},
    xtick=data,
    x tick label style={rotate=45, anchor=east},
    ytick={-0.2,-0.1,...,1},
    %legend pos=north east,
    legend pos=south east,
    legend style={at={(1.4,0.6)}},
    ymajorgrids=true,
    grid style=dashed,
    legend style={nodes={scale=0.8, transform shape}},
    every axis plot/.append style={ultra thick},
    %no markers,
    legend image post style={mark=*},
    cycle list name=exotic,
    mark=*,
    mark options={scale=0.6}
]
\addplot +[green]
    coordinates {
    (TREC5,0.511)(TREC6,0.689)(TREC7,0.778)(TREC8,0.822)(R04,0.733)(TREC\textsubscript{All},0.778)
    };
\addplot +[pink]
    coordinates {
    (TREC5,0.75)(TREC6,0.854)(TREC7,0.854)(TREC8,0.75)(R04,0.854)(TREC\textsubscript{All},0.854)
    };
    % nDCG
%\addplot +[blue]
%    coordinates {
%    (TREC5,0.644)(TREC6,0.911)(TREC7,0.822)(TREC8,0.778)(R04,0.867)(TREC\textsubscript{All},0.867)
 %   };
    \addplot +[red]
    coordinates {
    (TREC5,0.767)(TREC6,1.0)(TREC7,0.86)(TREC8,0.767)(R04,0.953)(TREC\textsubscript{All},0.953)
    };
    \addplot +[orange]
    coordinates {
    (TREC5,-0.095)(TREC6,0.23)(TREC7,0.322)(TREC8,0.23)(R04,0.296)(TREC\textsubscript{All},0.23)
    };

    \addplot +[dashed, green]
    coordinates {
    (TREC5,0.511)(TREC6,0.689)(TREC7,0.778)(TREC8,0.822)(R04,0.733)(TREC\textsubscript{All},0.778)
    };
\addplot +[dashed, pink]
    coordinates {
    (TREC5,0.759)(TREC6,0.887)(TREC7,0.887)(TREC8,0.782)(R04,0.887)(TREC\textsubscript{All},0.887)
    };
    % nDCG
%\addplot +[dashed, blue]
%    coordinates {
%    (TREC5,0.689)(TREC6,0.956)(TREC7,0.867)(TREC8,0.822)(R04,0.911)(TREC\textsubscript{All},0.911)
%    };
    \addplot +[dashed, red]
    coordinates {
    (TREC5,0.767)(TREC6,1.0)(TREC7,0.86)(TREC8,0.767)(R04,0.953)(TREC\textsubscript{All},0.953)
    };
    \addplot +[dashed, orange]
    coordinates {
    (TREC5,0.047)(TREC6,0.341)(TREC7,0.295)(TREC8,0.295)(R04,0.315)(TREC\textsubscript{All},0.25)
    };

    \legend{MAP,Rprec,P@100,P@10}
\end{axis}
\end{tikzpicture}
\caption{Kendall's $\tau$ scores for the system rankings compared to TREC collections when we use TNR documents and do not use TNR documents.  While solid lines denote the case when TNR documents are included, i.e., the original \testcol{}, dashed lines correspond to when TNR documents are removed, i.e, \testcol\textsubscript{wo\_TNR}.}
\label{fig:chatgpt23_vs_trec_wo_tnr}
\end{figure}
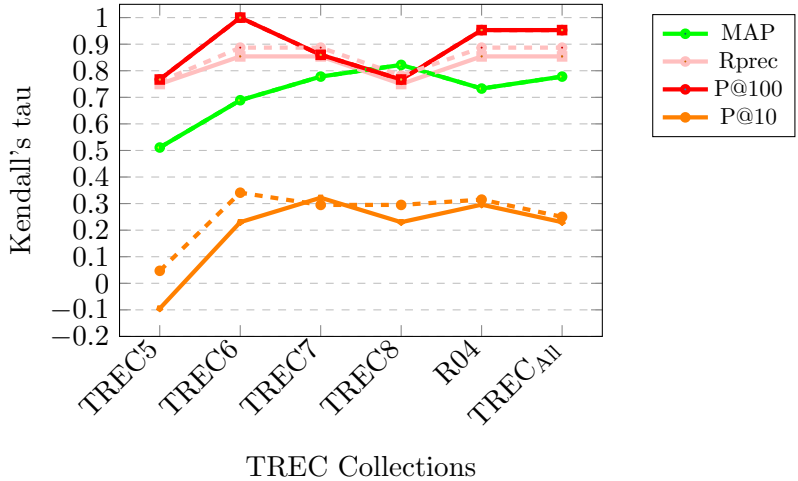

%In \textbf{Section \ref{sec_generate_nonrelevant_docs}}, we explain the reason and method of generating tricky non-relevant (TNR) documents. Since relevant documents in ChatGPT23 might be easy to find for IR systems, TNR documents make retrieval more complicated, containing common terms with relevant documents. Therefore, TNR documents play an essential role in better and more realistic evaluation. 
In this experiment, we examine the impact of  including tricky non-relevant (TNR) documents in \testcol{}.  
In particular, we exclude TNR documents from \testcol{} (\testcol\textsubscript{wo\_TNR}) and evaluate the IR systems accordingly. Next we compare the system rankings against TREC collections.   
\textbf{Figure \ref{fig:chatgpt23_vs_trec_wo_tnr}} shows the ranking correlation scores between \testcol\textsubscript{wo\_TNR} and TREC collections for four metrics.  We observe that exclusion of TNR documents has no impact rankings for MAP and P@100 metrics. However, in contrary to our expectations, the rank correlations increases in almost all cases for P@10 and RPrec. This might be because of challenges in creating TNR documents and possible errors in generation such as generating relevant documents instead of non-relevant ones.

%For MAP and P@100, using or not TNR does not affect ranking correlation, which results in overlapping solid and dashed lines. For the remaining metrics, including TNR documents decreases the ranking similarity between ChatGPT23 and TREC collections. 

\begin{comment}
\begin{table}[!htb]
\centering
  \begin{tabular}{|c|c|c|c|c|c|}
  \hline
    Collection & map & Rprec & ndcg & P.100 & P.10 \\
    \hline
    \hline
    chatgpt23 & 0.71 & 0.81 & 0.8 & 0.87 & 0.2 \\
    \hline
    trec5 & 0.53 & 0.76 & 0.78 & 0.83 & 0.65 \\
    \hline
    trec6 & 0.64 & 0.8 & 0.87 & 0.84 & 0.78 \\
    \hline
    trec7 & 0.73 & 0.83 & 0.87 & 0.87 & 0.78 \\
    \hline
    trec8 & 0.7 & 0.84 & 0.88 & 0.85 & 0.82 \\
    \hline
    robust2004 & 0.74 & 0.87 & 0.9 & 0.87 & 0.82 \\
    \hline
    combined & 0.8 & 0.9 & 0.92 & 0.9 & 0.86 \\
    \hline
    chatgpt23 (w/o TNR) & 0.71 & 0.84 & 0.85 & 0.87 & 0.26 \\
    \hline
  \end{tabular}
  \caption{Impact of Tricky Non-relevant Documents} 
  \label{tab_tnr_impact}
\end{table}
\end{comment}

%\input{figures/trec_correlation_tabs}
%\hl{uncomment the figure}
%\hl{the figure moved to the appendices}

\subsubsection{Impact of Randomly Generated Documents} \label{sec:impact_random}
\begin{figure}[!htb]
\centering
\begin{tikzpicture}
\begin{axis}[
    scale= 1,
    width=8cm,
    height=7cm,
    xlabel={TREC Collections},
    ylabel={Kendall's tau},
    ymin=-0.4, ymax=1.05,
    symbolic x coords={TREC5, TREC6, TREC7, TREC8, R04, TREC\textsubscript{All}},
    xtick=data,
    x tick label style={rotate=45, anchor=east},
    ytick={-0.3,-0.2,...,1},
    %legend pos=north east,
    legend pos=south east,
     legend style={at={(1.4,0.65)}},
    ymajorgrids=true,
    grid style=dashed,
    legend style={nodes={scale=0.8, transform shape}},
    every axis plot/.append style={ultra thick},
    %no markers,
    legend image post style={mark=*},
    cycle list name=exotic,
    mark=*,
    mark options={scale=0.6}
]
\addplot +[green]
    coordinates {
    (TREC5,0.511)(TREC6,0.689)(TREC7,0.778)(TREC8,0.822)(R04,0.733)(TREC\textsubscript{All},0.778)
    };
\addplot +[pink]
    coordinates {
    (TREC5,0.75)(TREC6,0.854)(TREC7,0.854)(TREC8,0.75)(R04,0.854)(TREC\textsubscript{All},0.854)
    };
    % ndcg
%\addplot +[blue]
%    coordinates {
 %   (TREC5,0.644)(TREC6,0.911)(TREC7,0.822)(TREC8,0.778)(R04,0.867)(TREC\textsubscript{All},0.867)
 %   };
    \addplot +[red]
    coordinates {
    (TREC5,0.767)(TREC6,1.0)(TREC7,0.86)(TREC8,0.767)(R04,0.953)(TREC\textsubscript{All},0.953)
    };
    \addplot +[orange]
    coordinates {
    (TREC5,-0.095)(TREC6,0.23)(TREC7,0.322)(TREC8,0.23)(R04,0.296)(TREC\textsubscript{All},0.23)
    };

    \addplot +[dashed, green]
    coordinates {
    (TREC5,0.511)(TREC6,0.689)(TREC7,0.778)(TREC8,0.822)(R04,0.733)(TREC\textsubscript{All},0.778)
    };
\addplot +[dashed, pink]
    coordinates {
    (TREC5,0.75)(TREC6,0.854)(TREC7,0.854)(TREC8,0.75)(R04,0.854)(TREC\textsubscript{All},0.854)
    };
%ndcg
%\addplot +[dashed, blue]
%    coordinates {
%    (TREC5,0.644)(TREC6,0.911)(TREC7,0.822)(TREC8,0.778)(R04,0.867)(TREC\textsubscript{All},0.867)
%    };
    \addplot +[dashed, red]
    coordinates {
    (TREC5,0.767)(TREC6,1.0)(TREC7,0.86)(TREC8,0.767)(R04,0.953)(TREC\textsubscript{All},0.953)
    };
    \addplot +[dashed, orange]
    coordinates {
    (TREC5,-0.306)(TREC6,-0.023)(TREC7,0.068)(TREC8,-0.023)(R04,0.045)(TREC\textsubscript{All},-0.023)
    };

    \legend{MAP,Rprec,P@100,P@10}
\end{axis}
\end{tikzpicture}
\caption{Kendall's $\tau$ of system ranking with respect to TREC collections when we use randomly generated documents (solid lines) and when we exclude them (dashed lines).  }
\label{fig:chatgpt23_vs_trec_wo_random}
\end{figure}
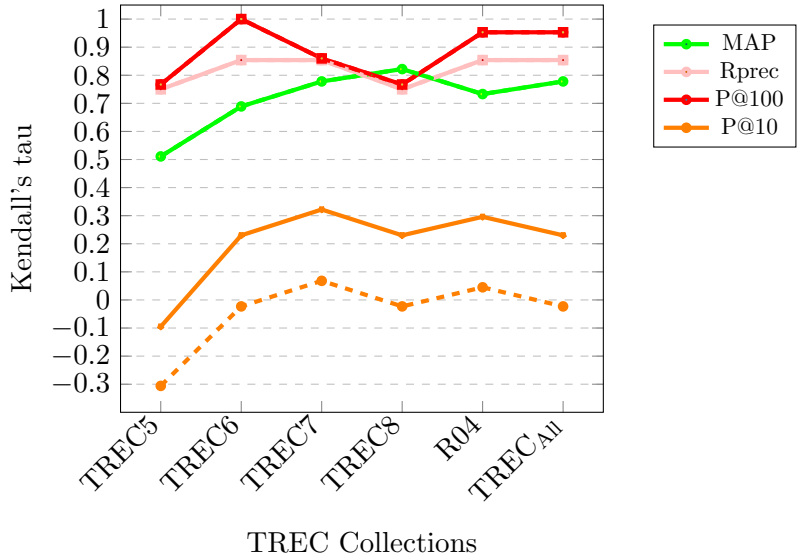

In this experiment, we explore the impact of randomly generated documents on the system ranking. Therefore, we first exclude all randomly generated documents from \testcol{} and then rank the IR systems based on the four metrics separately. Next, we calculate Kendall's $\tau$  correlation between the rankings we achieved with the rankings based on TREC collections. \textbf{Figure \ref{fig:chatgpt23_vs_trec_wo_random}} provides Kendall's $\tau$ scores for both cases. % when we exclude randomly generated documents and and also when we use all documents for easy comparison of their  impact on ranking. 
Interestingly, we observe that the rankings do not change for our metrics except P@10. Upon our deeper investigation, we found that the performance of IR systems are slightly affected in these metrics (i.e., MAP, RPrec, and P@100) because systems  do not retrieve the majority of them. Thus, the changes in performance scores do not cause a ranking change. Regarding P@10 metric, randomly generated documents have positive impact on system ranking.  

\subsubsection{Impact of Document Collection Size on System Ranking} \label{exp:sampled_disk45}
%doküman boyutlarını benzer tutuncaki sonuçlar

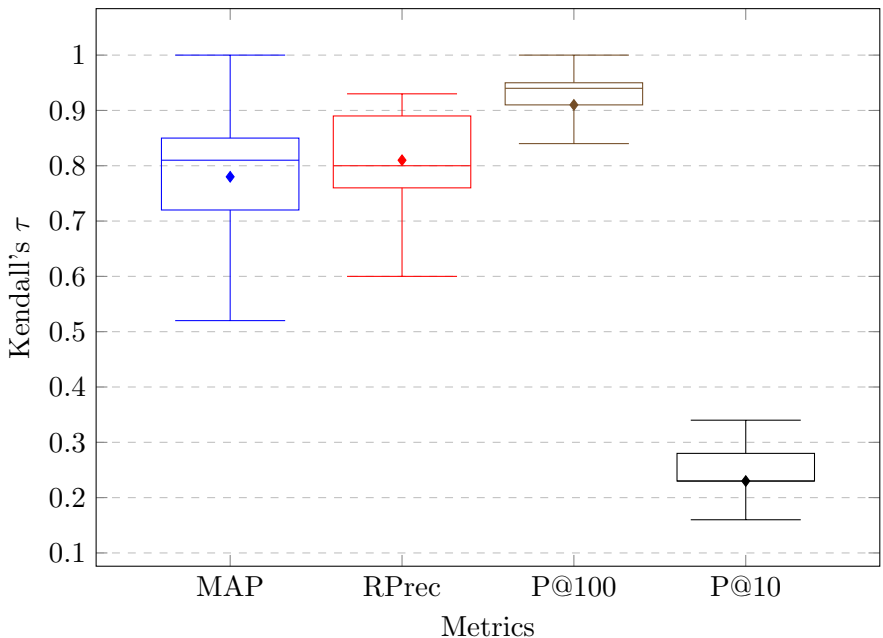
\begin{figure}[!htb]
\centering
\begin{tikzpicture}
  \begin{axis}
    [
    width=12cm,
    height=9cm,
    boxplot/draw direction = y,
    xlabel={Metrics},
    ylabel={ Kendall's $\tau$},
    xtick={1,2,3,4},
    xticklabels={MAP, RPrec, P@100, P@10},
    ytick={0,0.1,...,1},
    ymajorgrids=true,
    grid style=dashed,
    ]
    \addplot+[
    boxplot prepared={
      median=0.81,
      upper quartile=0.85,
    lower quartile=0.72,
    upper whisker=1.0,
    lower whisker=0.52,
    average=0.78
    },
    ] coordinates {};
    \addplot+[
    boxplot prepared={
      median=0.8,
upper quartile=0.89,
lower quartile=0.76,
upper whisker=0.93,
lower whisker=0.6,
average=0.81
    },
    ] coordinates {};
    \addplot+[
    boxplot prepared={
      median=0.94,
upper quartile=0.95,
lower quartile=0.91,
upper whisker=1.0,
lower whisker=0.84,
average=0.91
    },
    ] coordinates {};
    \addplot+[
    boxplot prepared={
      median=0.23,
upper quartile=0.28,
lower quartile=0.23,
upper whisker=0.34,
lower whisker=0.16,
average=0.23
    },
    ] coordinates {};
  \end{axis}
\end{tikzpicture}
\caption{Box plot representation of Kendall's $\tau$ scores between rankings when we use all documents in Disks 4-5 vs. a sample of documents. The diamonds represent the average $\tau$ score.  Each sample contains randomly picked 96,196 documents, matching the size of \testcol{}. We use all available topics in TREC collections we have (i.e., TREC\textsubscript{All}).}
\label{fig:sampled_disk45_box}
\end{figure}

One of the most apparent differences between \testcol{} and TREC’s Disk 4–5 collection is that \testcol{} contains roughly 5.5 times fewer documents, which may significantly affect system rankings. To investigate this effect, we create smaller versions of Disk 4–5 by randomly sampling 96,196 documents, matching the size of \testcol{}. After forming 100 such samples, we rank the systems based on each sample and measure the correlation between these rankings and the ranking obtained when using all documents in Disk 4–5. For this evaluation, we use the full set of available TREC topics, i.e., TREC\textsubscript{All}. \textbf{Figure \ref{fig:sampled_disk45_box}} shows the resulting Kendall’s $\tau$ scores for each metric as a box plot.

%One of the most apparent differences of \testcol{} from  TREC's document collection, i.e., Disk 4-5, is that it consists of almost 5.5 times fewer documents, which might have a huge impact on system rankings. To investigate this, we create smaller versions of Disk4-5, by sampling 96,196 documents, i.e., the size of \testcol. After forming 100 samples by random sampling, we rank systems based on each sample and measure ranking correlation between the system ranking when we use all documents in Disk4-5. We use all available search topics in TREC collections we have, i.e., TREC\textsubscript{All}. \textbf{Figure \ref{fig:sampled_disk45_box}} shows Kendall's $\tau$ scores we obtained for each metric as a box plot. 

%average Kendall's $\tau$ correlation between Disk4-5 samples and complete Disk4-5 for each TREC collection. 

%\begin{figure}[!htb]
%\centering
%\input{figures/sampled_disk45}
%\caption{doküman boyutlarını benzer tutunca TREC koleksiyonlarının kendisiyle olan korelasyonları}
%\label{fig:sampled_disk45}
%\end{figure} 

As shown in the figure, the ranking correlation between \testcol{} and TREC\textsubscript{All} is similar to the correlations we observe when  the number of documents in Disk 4–5 is reduced. In particular, \testcol{} achieves Kendall’s $\tau$ scores of 0.78, 0.85, 0.95, and 0.23 for MAP, RPrec, P@100, and P@10, respectively (see Table \ref{tab_tau_map_rprec_prec}). By comparison, when using TREC\textsubscript{All} on the full Disk 4–5 collection versus the 96,196-document samples, the average Kendall’s $\tau$ scores are 0.78, 0.81, 0.91, and 0.23 for these same metrics. Overall, these results show that collection size can substantially affect system rankings. %They also imply that we may achieve rankings more similar to those observed with TREC collections if we generate as many documents as Disks 4-5.

\subsubsection{Impact of Prevalence of Relevant Documents} \label{exp:prevalence_rel_docs}
%ilgili doküman sayısı değiştikçe nasıl sonuçlar değişiyor

\begin{figure}[!htb]
\centering
\begin{tikzpicture}
\begin{axis}[
    scale= 1,
    width=12cm,
    height=9cm,
    xlabel={ $N_R$},
    ylabel={Kendall's $\tau$},
    ymin=0, ymax=1.05,
    %symbolic x coords={trec5, trec6, trec7, trec8, robust2004, combined},
    %xtick=data,
    %x tick label style={rotate=45, anchor=east},
    xtick={10,20,...,90},
    ytick={0, 0.1,...,1},
    %legend pos=north east,
    %legend pos=north,
    ymajorgrids=true,
    legend columns=-1,
    grid style=dashed,
    legend style={nodes={scale=0.8, transform shape}},
    every axis plot/.append style={ultra thick},
    %no markers,
    legend image post style={mark=*},
    cycle list name=exotic,
    mark=*,
    mark options={scale=0.6}
]
\addplot +[green]
    coordinates {
    (10,0.636)(20,0.707)(30,0.751)(40,0.751)(50,0.707)(60,0.707)(70,0.707)(80,0.707)(90,0.707)
    };
\addplot +[pink]
    coordinates {
    (10,0.753)(20,0.812)(30,0.812)(40,0.812)(50,0.812)(60,0.812)(70,0.812)(80,0.812)(90,0.812)
    };
    % ndcg
%\addplot +[blue]
%    coordinates {
%    (10,0.71)(20,0.76)(30,0.804)(40,0.836)(50,0.804)(60,0.804)(70,0.804)(80,0.804)(90,0.804)
 %   };
    \addplot +[red]
    coordinates {
    (10,0.816)(20,0.87)(30,0.87)(40,0.87)(50,0.87)(60,0.87)(70,0.87)(80,0.87)(90,0.87)
    };
    \addplot +[orange]
    coordinates {
    (10,0.267)(20,0.152)(30,0.134)(40,0.19)(50,0.272)(60,0.16)(70,0.171)(80,0.171)(90,0.175)
    };
    \legend{MAP,Rprec,P@100,P@10}
\end{axis}
\end{tikzpicture}
\caption{ Average Kendall's $\tau$ score between system rankings based on \testcol{} and TREC collections for varying number of relevant documents per topic in \testcol{}. $N_R$ represents the maximum number of relevant documents per topic.}
\label{fig:impact_of_number_of_relevant_docs}
\end{figure}
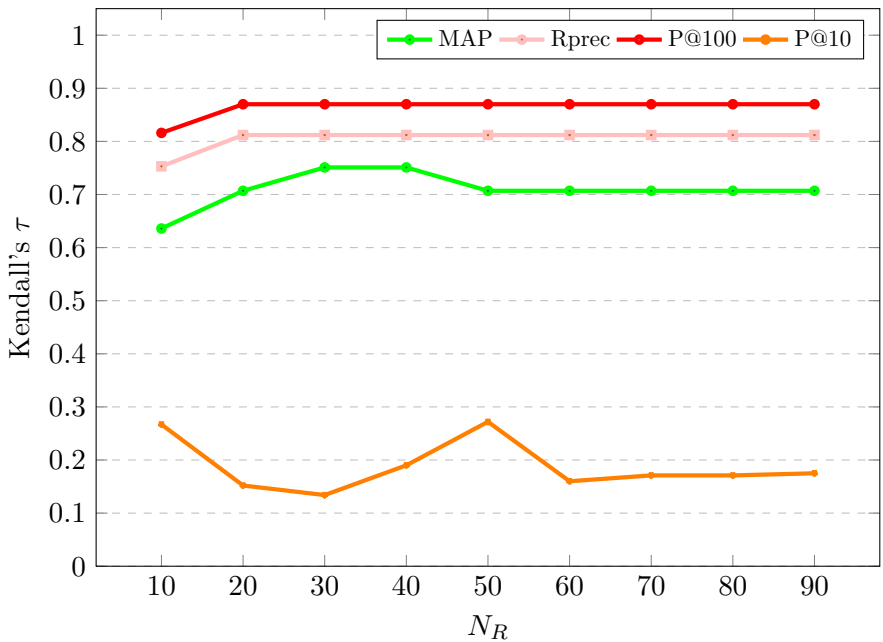

%While generating relevant documents, we employ ChatGPT to generate 100 documents for each topic (though, ChatGPT may not generate exactly 100 documents for a topic). 
In this experiment, we explore how the evaluation of systems are affected with varying number of relevant documents per topic. In particular, we vary the maximum number of relevant documents per topic ($N_R$) in \testcol{} from 10 to 100 and rank systems accordingly. Next, we compare system rankings with TREC collections. 
%evaluate system using \testcol{} but changing the number of relevant documents generated per topic  from 10 to 100. Next, we calculate Kendall's $\tau$ score based on the system rankings based on TREC collections.  %For example, we select the first X  ($X \in [10 - 100]$) relevant documents, ignore the rest and carry out the evaluation with top X relevant document for each topic.
%Then, we measure the ranking correlation between ChatGPT23 and TREC collections.
\textbf{Figure \ref{fig:impact_of_number_of_relevant_docs}} shows how the average tau correlations across TREC collections for  varying maximum number of relevant topics per topic.

We observe that  $\tau$ score increases when $N_R$ increases from 10 to 20, and then become stable for P@100 and Rprec metrics.   Regarding MAP score, the maximum  $\tau$ score is achieved with $N_R=30$. The ranking correlation becomes stabilized for $N_R \geq 50$. The results for P@10 only becomes stabilized after $N_R \geq 60$ and are affected by the number of relevant documents per topic more than the other metrics. Overall, varying the maximum number of relevant documents per topic does not produce any results that would alter the conclusions drawn from our previous experiments.

%\subsubsection{Impact of Topics}

\section{Limitations} \label{sec:discussion_limitations}

In this section, we discuss the limitations of our study and draw a roadmap for future studies.

\subsection{Document Collection}
% more documents.

The size of a document collection is vital for the reliability of a test collection. In our proof-of-concept study, we employed ChatGPT to generate 96,196 documents and explored various prompts to increase textual diversity. Despite consisting of a substantial number of documents, \testcol{} remains relatively small compared to widely used collections such as Disks 4–5, which contain around 500,000 documents, and web-based collections that are even larger. Conducting experiments with a larger set of documents might yield different outcomes. To address this limitation, we illustrate the impact of collection size on system rankings in Section \ref{exp:sampled_disk45}; however, we leave the exploration of larger document sets for future work.
%The size of a document collection is vital for the reliability of a test collection. In our proof-of-concept study, we employed ChatGPT to generate 96,196 documents and explored various prompts to increase textual diversity. Despite consisting of many documents, \testcol{} is not a large test collection compared to other popular collections. For instance, our baseline document collection, Disks 4-5, contains around 500K documents and the web collection contains even much more documents. Conducting experiments with a larger set of documents might result in different outcomes. To reduce the impact of this limitation, we show how the collection size can affect system rankings in Section \ref{exp:sampled_disk45}. Nevertheless, we leave conducting experiments with larger document collection as future work. 
%Consequently, to be an alternative to traditional test collections, ChatGPT23 needs to have more documents. 

The diversity of the generated documents is another key factor in evaluating systems reliably. In this study, we exclusively used ChatGPT with a fixed set of prompts. However, the content of generated documents may be influenced by various factors, such as the choice of the language model and the prompts used. Exploring other large language models (e.g., LLAMA \cite{touvron2023llama}) to generate documents represents an interesting direction for future research.
%The diversity of the generated documents is another important factor in evaluating systems reliably. In our study, we exclusively used ChatGPT with a specific set of prompts. However, the content of generated documents can be influenced by various factors, such as the choice of the language model and the prompts used. We believe that exploring different LLMs to generate documents such as LLAMA \cite{touvron2023llama} and Gemini is an interesting research direction.

Regarding the prompt design, we examined various prompts for document generation (Section \ref{sec:automatic_document_generation}). Initially, we instructed ChatGPT to produce news articles, because Disks 4-5 also consists of news articles. However, this approach yielded limited textual diversity, leading us to modify our prompt to generate long texts, which improved diversity. We also incorporated subtopics to further enhance textual diversity. Nevertheless, prompt selection undeniably influences generated content, so exploring alternative prompts could be highly beneficial for constructing test collections.
%Regarding the prompt design, we examined various prompts for document generation (Section \ref{sec:automatic_document_generation}). In particular, we initially utilized those derived from our first set of news articles and subsequently created several other variations.  For instance, our first approach was to prompt ChatGPT to generate news articles because Disks 4-5 also consists of news articles. However, the results were not satisfactory because of limited textual diversity. Therefore, we modified our prompt to generate long texts, which yields a better result. We also investigate subtopics to increase textual diversity. However, the choice of prompts undeniably impacts the generated content. Thus, it is worth exploring other prompts for test collection construction. 

Lastly, we did not explore generating documents in different file formats, as our comparison focuses on TREC collections that  contain only news articles. We leave the creation of test collections using other content types, such as web pages and scholarly articles, as future work.

\subsection{Topics}

In our work, we use topics from existing collections to generate documents. This design choice allows us to make a fair comparison against established test collections. However, we could also generate topics using LLMs, as explored in  prior work \cite{rahmani2024synthetic}. Such an approach would enable fully automated test collection construction without relying on any resource other than LLMs. We leave this work as future work because it makes the evaluation of the resulting collection more challenging. Moreover, having control over topic selection allows us to control the evaluation process with minimal manual effort.
%In our work, we use existing topic collections to generate documents. This design choice allows us to make a fair comparison against real test collections. However, we can also generate topics using LLMs, as several studies in the literature have already investigated that \textbf{[REF]}. This will also enable generating  test collections fully automatic without relying on any resource other than LLMs. We leave this work as future work because it makes evaluation of the collection more challenging. Furthermore, deciding which topics allow us to control the evaluation process with a minimal manual effort. 

\subsection{Relevance Judgments}

In our approach, we assume that documents generated for a specific topic are relevant to that topic, while documents generated for other topics are considered non-relevant. Although this assumption may not always hold, it aligns with the standard pooling technique, where unjudged documents are treated as non-relevant.   In our study, we specifically investigate the impact of this assumption  by conducting multiple experiments, including an analysis of randomly generated documents. Our results show that evaluations conducted using \testcol{} are comparable to those based on existing TREC collections when assessed with P@100, MAP, and RPrec metrics. However, our analysis reveals that our assumption does not hold for all documents. Future work can focus on prompting strategies  to improve the accuracy of generating relevant documents.   Such improvements in accuracy are likely to lead to more reliable and robust system evaluations.
%In our approach, we assume that the documents generated for a particular topic are relevant to that topic while the other documents are non-relevant. Obviously, this assumption might not hold in all cases. However, note that a similar assumption also exist in standard pooling technique such that not-judged documents are assumed to be non-relevant. Furthermore, in our experiments, we also evaluate the impact of documents randomly generated. \todo{belki buraya sonuclardan sonra bir sey eklenebilir}

%After being proposed by \cite{spark1975report}, pooling has been adopted as a standard procedure for reducing the cost of test collection creation.
%While cost-effective, it relies on the assumption that documents out of the pool are irrelevant.Nevertheless, pooling relies on the assumption that documents out of the pool are irrelevant. In our method, we similarly assume that the documents generated for all other topics are non-relevant to a topic without judgment. 

Another limitation of our methodology is that it relies on binary relevance judgments, without differentiating levels of document relevance. LLMs could potentially be used to predict the grade of relevance or generate documents at varying relevance levels, which is a direction we leave for future investigation.
%Moreover, in our methodology, we do not have any graded relevance judgments. Instead we rely on binary judgments. LLMs might again be used to predict the grade of the relevance or they can be prompted to generate documents at different relevance levels. We leave this study as future work. 

\subsection{Evaluation}
%systems might have been more diverse and more.

Assessing the reliability of our test collection requires the design and implementation of IR systems capable of operating on it. This process involves several critical considerations. For example, a diverse and sufficiently large set of systems is essential to evaluate whether the collection can effectively differentiate them. However, developing such an extensive set of IR systems is costly. In this study, we could implement 10 IR systems with different retrieval strategies.  The exploration of a more extensive set is left for future research.

%In order to evaluation the evaluation reliability of our collection, we need IR systems that can run on our collection. However, there are several things that needs to be considered in this step. For instance, it is important  to use a high number of systems. In addition, the systems we use should cover a wide range of retrieval methods which  have different retrieval performance to assess whether our collection can distinguish these different systems. However, developing such a set of IR systems is extremely costly. In our study, we could develop  10 IR systems for our evaluations, covering various retrieval strategies. We leave employing a larger set of IR systems as future work. 

Moreover, determining the reliability of an evaluation method is inherently challenging, given that we cannot definitively know which IR system is ``truly" better. Therefore, to evaluate our test collection, we  analyze evaluation stability and compare its system rankings with those of existing test collections, under the assumption that strong correlations imply similar evaluation quality. However, it is important to note that even perfect correlation does not ensure validity in all cases. To address this limitation, we adopt established methodologies from the literature to assess evaluation reliability and use well-known high-quality TREC collections such as TREC-8 \cite{voorhees2018building}. Voorhees et al. \cite{voorhees2022can} report that TREC-8 remains reliable for assessing modern neural retrieval models despite being created decades ago. However, exploring additional TREC collections —particularly more recent ones— remains an important direction. Since such assessments require different documents and topics, we leave this exploration for future work.

\section{Conclusion} \label{sec:conclusion}

In this study, we introduce a new approach to creating IR test collections using LLMs. In particular, our approach relies on the assumption that documents generated by an LLM are naturally relevant to the prompts they are based on. Utilizing this principle, we used existing TREC search topics to generate  documents. Each document is deemed relevant only to its specific generating prompt, while other document-topic pairs are treated as non-relevant. Furthermore, we generated non-relevant documents to introduce realistic challenges for retrieval systems.
%and present \testcol{}, the first LLM-generated test collection to evaluate IR systems. %Our goal is to reduce the high cost associated with manual relevance judgments.
%In particular, through prompt engineering, we devised a method for generating relevant, ``tricky'' non-relevant, and random documents. %In \ref{sec:relevance_analysis}, we justify the validity of this method. 
Using these techniques, we created \testcol{} with 96,196 documents, 300 topics, and 18,964 relevant document-topic pairs.  

%and then carried out a thorough analysis, comparing it to established TREC collections.

%As a result of prompt engineering, we come up with an approach. Our approach includes generating relevant, tricky non-relevant and randomly generated documents. In Section \ref{sec:relevance_analysis}, we justify the validity of our method. %In Section \ref{sec:manual_subtopics}, we demonstrate how easy utilizing ChatGPT makes subtopic generation instead of manually creating them.

In our comprehensive experiments, we first analyze both the quantitative and linguistic properties of our collection. Our analysis reveals the following key findings: \testcol{} includes shorter documents and sentences with lower lexical diversity, while requiring a higher educational level compared to the documents in Disks 4-5. However, the topical diversity per document is similar in both collections.

Next, we evaluate the system rankings generated by \testcol{} by comparing them with those derived from TREC collections. Our findings indicate that system rankings based on \testcol{} align closely with those from TREC collections when evaluated using P@100, MAP, and RPrec metrics. However, rankings differ significantly when using the P@10 metric. %P@10 showed the lowest Kendall’s $\tau$ correlation values, while P@100 yielded the highest correlations. 
Moreover, we examine the impact of tricky non-relevant and random documents on ranking correlations. Contrary to our expectations, tricky non-relevant documents either have no effect or slightly decrease ranking correlations. On the other hand, random documents increase correlations for P@10 but do not affect the system rankings for other metrics. Finally, we observe that variations in collection size alone can substantially alter system rankings. This suggests that if \testcol{} were scaled to include a comparable number of documents as Disks 4-5, it might produce rankings more closely aligned with TREC test collections.

 Our work can be expanded in several directions for future research. First, we aim to investigate the generation of other document types, such as web pages and social media posts. Test collections that contain generated social media posts are particularly valuable due to the inherent challenges of collecting such data, including the dynamic nature of platforms, redistribution data policies of platforms, and privacy concerns. In addition, we plan to generate documents with graded relevance levels, enabling more nuanced evaluation scenarios. Finally, expanding both the number of generated documents and the diversity of IR systems included in the experiments will allow us to evaluate our approach in contexts that more closely resemble real-world scenarios.

\bibliographystyle{acm}
\bibliography{references}

\begin{thebibliography}{10}

\bibitem{altun2020building}
{\sc Altun, B., and Kutlu, M.}
\newblock Building test collections using bandit techniques: a reproducibility
  study.
\newblock In {\em Proceedings of the 29th ACM International Conference on
  Information \& Knowledge Management\/} (2020), pp.~1953--1956.

\bibitem{asadi2011pseudo}
{\sc Asadi, N., Metzler, D., Elsayed, T., and Lin, J.}
\newblock Pseudo test collections for learning web search ranking functions.
\newblock In {\em Proceedings of the 34th international ACM SIGIR conference on
  Research and development in Information Retrieval\/} (2011), pp.~1073--1082.

\bibitem{askari2023test}
{\sc Askari, A., Aliannejadi, M., Kanoulas, E., and Verberne, S.}
\newblock A test collection of synthetic documents for training rankers:
  Chatgpt vs. human experts.
\newblock In {\em Proceedings of the 32nd ACM International Conference on
  Information and Knowledge Management\/} (2023), pp.~5311--5315.

\bibitem{askari2023expand}
{\sc Askari, A., Aliannejadi, M., Meng, C., Kanoulas, E., and Verberne, S.}
\newblock Expand, highlight, generate: Rl-driven document generation for
  passage reranking.
\newblock In {\em Proceedings of the 2023 Conference on Empirical Methods in
  Natural Language Processing\/} (2023), pp.~10087--10099.

\bibitem{aslam2007inferring}
{\sc Aslam, J.~A., and Yilmaz, E.}
\newblock Inferring document relevance from incomplete information.
\newblock In {\em Proceedings of the sixteenth ACM conference on Conference on
  information and knowledge management\/} (2007), pp.~633--642.

\bibitem{azzopardi2006automatic}
{\sc Azzopardi, L., and De~Rijke, M.}
\newblock Automatic construction of known-item finding test beds.
\newblock In {\em Proceedings of the 29th annual international ACM SIGIR
  conference on Research and Development in Information Retrieval\/} (2006),
  pp.~603--604.

\bibitem{azzopardi2007building}
{\sc Azzopardi, L., De~Rijke, M., and Balog, K.}
\newblock Building simulated queries for known-item topics: an analysis using
  six european languages.
\newblock In {\em Proceedings of the 30th annual international ACM SIGIR
  conference on Research and development in information retrieval\/} (2007),
  pp.~455--462.

\bibitem{benedict2024gen}
{\sc B{\'e}n{\'e}dict, G., Zhang, R., Metzler, D., Yates, A., and Jiang, Z.}
\newblock Gen-ir@ sigir 2024: The second workshop on generative information
  retrieval.
\newblock In {\em Proceedings of the 47th International ACM SIGIR Conference on
  Research and Development in Information Retrieval\/} (2024), pp.~3029--3032.

\bibitem{berendsen2012generating}
{\sc Berendsen, R., Tsagkias, M., De~Rijke, M., and Meij, E.}
\newblock Generating pseudo test collections for learning to rank scientific
  articles.
\newblock In {\em Information Access Evaluation. Multilinguality,
  Multimodality, and Visual Analytics: Third International Conference of the
  CLEF Initiative, CLEF 2012, Rome, Italy, September 17-20, 2012. Proceedings
  3\/} (2012), Springer, pp.~42--53.

\bibitem{berendsen2013pseudo}
{\sc Berendsen, R., Tsagkias, M., Weerkamp, W., and De~Rijke, M.}
\newblock Pseudo test collections for training and tuning microblog rankers.
\newblock In {\em Proceedings of the 36th international ACM SIGIR conference on
  Research and development in information retrieval\/} (2013), pp.~53--62.

\bibitem{buckley2004retrieval}
{\sc Buckley, C., and Voorhees, E.~M.}
\newblock Retrieval evaluation with incomplete information.
\newblock In {\em Proceedings of the 27th annual international ACM SIGIR
  conference on Research and development in information retrieval\/} (2004),
  pp.~25--32.

\bibitem{bueno2024quati}
{\sc Bueno, M., de~Oliveira, E.~S., Nogueira, R., Lotufo, R.~A., and Pereira,
  J.~A.}
\newblock Quati: A brazilian portuguese information retrieval dataset from
  native speakers.
\newblock {\em arXiv preprint arXiv:2404.06976\/} (2024).

\bibitem{buttcher2007reliable}
{\sc B{\"u}ttcher, S., Clarke, C.~L., Yeung, P.~C., and Soboroff, I.}
\newblock Reliable information retrieval evaluation with incomplete and biased
  judgements.
\newblock In {\em Proceedings of the 30th annual international ACM SIGIR
  conference on Research and development in information retrieval\/} (2007),
  pp.~63--70.

\bibitem{carterette2005incremental}
{\sc Carterette, B., and Allan, J.}
\newblock Incremental test collections.
\newblock In {\em Proceedings of the 14th ACM international conference on
  information and knowledge management\/} (2005), pp.~680--687.

\bibitem{chen2024genqa}
{\sc Chen, J., Qadri, R., Wen, Y., Jain, N., Kirchenbauer, J., Zhou, T., and
  Goldstein, T.}
\newblock Genqa: Generating millions of instructions from a handful of prompts.
\newblock {\em arXiv preprint arXiv:2406.10323\/} (2024).

\bibitem{wt14}
{\sc Collins-Thompson, K., Macdonald, C., Bennett, P., Diaz, F., and Voorhees,
  E.~M.}
\newblock Trec 2014 web track overview.
\newblock Tech. rep., MICHIGAN UNIV ANN ARBOR, 2015.

\bibitem{cooper1973simulation}
{\sc Cooper, M.~D.}
\newblock A simulation model of an information retrieval system.
\newblock {\em Information Storage and Retrieval 9}, 1 (1973), 13--32.

\bibitem{cormack2018beyond}
{\sc Cormack, G.~V., and Grossman, M.~R.}
\newblock Beyond pooling.
\newblock In {\em The 41st International ACM SIGIR Conference on Research \&
  Development in Information Retrieval\/} (2018), pp.~1169--1172.

\bibitem{cormack1998efficient}
{\sc Cormack, G.~V., Palmer, C.~R., and Clarke, C.~L.}
\newblock Efficient construction of large test collections.
\newblock In {\em Proceedings of the 21st annual international ACM SIGIR
  conference on Research and development in information retrieval\/} (1998),
  pp.~282--289.

\bibitem{de2024exploring}
{\sc de~Jesus, G., and Nunes, S.}
\newblock Exploring large language models for relevance judgments in tetun.
\newblock {\em arXiv preprint arXiv:2406.07299\/} (2024).

\bibitem{dietz2022wikimarks}
{\sc Dietz, L., Chatterjee, S., Lennox, C., Kashyapi, S., Oza, P., and Gamari,
  B.}
\newblock Wikimarks: harvesting relevance benchmarks from wikipedia.
\newblock In {\em Proceedings of the 45th International ACM SIGIR Conference on
  Research and Development in Information Retrieval\/} (2022), pp.~3003--3012.

\bibitem{dietz2020humans}
{\sc Dietz, L., and Dalton, J.}
\newblock Humans optional? automatic large-scale test collections for entity,
  passage, and entity-passage retrieval.
\newblock {\em Datenbank-Spektrum 20}, 1 (2020), 17--28.

\bibitem{eguchi2002overview}
{\sc Eguchi, K., Oyama, K., Ishida, E., Kando, N., and Kuriyama, K.}
\newblock Overview of the web retrieval task at the third ntcir workshop.
\newblock In {\em NTCIR\/} (2002), Citeseer.

\bibitem{faggioli2023perspectives}
{\sc Faggioli, G., Dietz, L., Clarke, C.~L., Demartini, G., Hagen, M., Hauff,
  C., Kando, N., Kanoulas, E., Potthast, M., Stein, B., et~al.}
\newblock Perspectives on large language models for relevance judgment.
\newblock In {\em Proceedings of the 2023 ACM SIGIR International Conference on
  Theory of Information Retrieval\/} (2023), pp.~39--50.

\bibitem{farzi2024exam}
{\sc Farzi, N., and Dietz, L.}
\newblock An exam-based evaluation approach beyond traditional relevance
  judgments.
\newblock {\em arXiv preprint arXiv:2402.00309\/} (2024).

\bibitem{flesch1949art}
{\sc Flesch, R.~F., and Gould, A.~J.}
\newblock The art of readable writing.
\newblock {\em (No Title)\/} (1949).

\bibitem{grady2010crowdsourcing}
{\sc Grady, C., and Lease, M.}
\newblock Crowdsourcing document relevance assessment with mechanical turk.
\newblock In {\em Proceedings of the NAACL HLT 2010 workshop on creating speech
  and language data with Amazon’s mechanical turk\/} (2010), pp.~172--179.

\bibitem{grootendorst2022bertopic}
{\sc Grootendorst, M.}
\newblock Bertopic: Neural topic modeling with a class-based tf-idf procedure.
\newblock {\em arXiv preprint arXiv:2203.05794\/} (2022).

\bibitem{guiver2009few}
{\sc Guiver, J., Mizzaro, S., and Robertson, S.}
\newblock A few good topics: Experiments in topic set reduction for retrieval
  evaluation.
\newblock {\em ACM Transactions on Information Systems (TOIS) 27}, 4 (2009),
  1--26.

\bibitem{trec8-overview}
{\sc Harman, D.}
\newblock Overview of the eighth text retrieval conference.

\bibitem{harman1996overview}
{\sc Harman, D., and Voorhees, E.}
\newblock Overview of the fifth text retrieval conference (trec-5).
\newblock In {\em Information Technology: The Fifth Text REtrieval Conference
  (TREC-5), D. Harman and E. Voorhees, eds., National Institute of Standards
  and Technology Special Publication\/} (1996), pp.~500--238.

\bibitem{trec5}
{\sc Harman, D., and Voorhees, E.}
\newblock Overview of the fifth text retrieval conference (trec-5).
\newblock In {\em Information Technology: The Fifth Text REtrieval Conference
  (TREC-5), D. Harman and E. Voorhees, eds., National Institute of Standards
  and Technology Special Publication\/} (1996), pp.~500--238.

\bibitem{hasanain2020artest}
{\sc Hasanain, M., Barkallah, Y., Suwaileh, R., Kutlu, M., and Elsayed, T.}
\newblock Artest: The first test collection for arabic web search with
  relevance rationales.
\newblock In {\em Proceedings of the 43rd international ACM sigir conference on
  research and development in information retrieval\/} (2020), pp.~2017--2020.

\bibitem{hauff2010retrieval}
{\sc Hauff, C., and de~Jong, F.}
\newblock Retrieval system evaluation: Automatic evaluation versus incomplete
  judgments.
\newblock In {\em Proceedings of the 33rd international ACM SIGIR conference on
  Research and development in information retrieval\/} (2010), pp.~863--864.

\bibitem{hawking2020simulating}
{\sc Hawking, D., Billerbeck, B., Thomas, P., and Craswell, N.}
\newblock {\em Simulating information retrieval test collections}.
\newblock Springer, 2020.

\bibitem{hosseini2012uncertainty}
{\sc Hosseini, M., Cox, I.~J., Milic-Frayling, N., Shokouhi, M., and Yilmaz,
  E.}
\newblock An uncertainty-aware query selection model for evaluation of ir
  systems.
\newblock In {\em Proceedings of the 35th international ACM SIGIR conference on
  Research and development in information retrieval\/} (2012), pp.~901--910.

\bibitem{jones1975report}
{\sc Jones, S.}
\newblock Report on the need for and provision of an" ideal" information
  retrieval test collection.

\bibitem{kanoulas2018clef}
{\sc Kanoulas, E., Li, D., Azzopardi, L., and Spijker, R.}
\newblock Clef 2018 technologically assisted reviews in empirical medicine
  overview.
\newblock In {\em CEUR workshop proceedings\/} (2018), vol.~2125.

\bibitem{kim2009retrieval}
{\sc Kim, J., and Croft, W.~B.}
\newblock Retrieval experiments using pseudo-desktop collections.
\newblock In {\em Proceedings of the 18th acm conference on information and
  Knowledge Management\/} (2009), pp.~1297--1306.

\bibitem{kincaid1975derivation}
{\sc Kincaid, J.~P., Fishburne~Jr, R.~P., Rogers, R.~L., and Chissom, B.~S.}
\newblock Derivation of new readability formulas (automated readability index,
  fog count and flesch reading ease formula) for navy enlisted personnel.

\bibitem{kutlu2018intelligent}
{\sc Kutlu, M., Elsayed, T., and Lease, M.}
\newblock Intelligent topic selection for low-cost information retrieval
  evaluation: A new perspective on deep vs. shallow judging.
\newblock {\em Information Processing \& Management 54}, 1 (2018), 37--59.

\bibitem{Lin_etal_SIGIR2021_Pyserini}
{\sc Lin, J., Ma, X., Lin, S.-C., Yang, J.-H., Pradeep, R., and Nogueira, R.}
\newblock {Pyserini}: A {Python} toolkit for reproducible information retrieval
  research with sparse and dense representations.
\newblock In {\em Proceedings of the 44th Annual International ACM SIGIR
  Conference on Research and Development in Information Retrieval (SIGIR
  2021)\/} (2021), pp.~2356--2362.

\bibitem{lin2023aggretriever}
{\sc Lin, S.-C., Li, M., and Lin, J.}
\newblock Aggretriever: A simple approach to aggregate textual representations
  for robust dense passage retrieval.
\newblock {\em Transactions of the Association for Computational Linguistics
  11\/} (2023), 436--452.

\bibitem{lin2021batch}
{\sc Lin, S.-C., Yang, J.-H., and Lin, J.}
\newblock In-batch negatives for knowledge distillation with tightly-coupled
  teachers for dense retrieval.
\newblock In {\em Proceedings of the 6th Workshop on Representation Learning
  for NLP (RepL4NLP-2021)\/} (2021), pp.~163--173.

\bibitem{macavaney2023one}
{\sc MacAvaney, S., and Soldaini, L.}
\newblock One-shot labeling for automatic relevance estimation.
\newblock In {\em Proceedings of the 46th International ACM SIGIR Conference on
  Research and Development in Information Retrieval\/} (2023), pp.~2230--2235.

\bibitem{mass1972zusammenhang}
{\sc Mass, H.-D.}
\newblock {\"U}ber den zusammenhang zwischen wortschatzumfang und l{\"a}nge
  eines textes.
\newblock {\em Zeitschrift f{\"u}r Literaturwissenschaft und Linguistik 2}, 8
  (1972), 73.

\bibitem{mccarthy2005assessment}
{\sc McCarthy, P.~M.}
\newblock {\em An assessment of the range and usefulness of lexical diversity
  measures and the potential of the measure of textual, lexical diversity
  (MTLD)}.
\newblock PhD thesis, The University of Memphis, 2005.

\bibitem{mccarthy2007vocd}
{\sc McCarthy, P.~M., and Jarvis, S.}
\newblock vocd: A theoretical and empirical evaluation.
\newblock {\em Language Testing 24}, 4 (2007), 459--488.

\bibitem{mcdonnell2016relevant}
{\sc McDonnell, T., Lease, M., Kutlu, M., and Elsayed, T.}
\newblock Why is that relevant? collecting annotator rationales for relevance
  judgments.
\newblock In {\em Proceedings of the AAAI Conference on Human Computation and
  Crowdsourcing\/} (2016), vol.~4.

\bibitem{mehrdad2024large}
{\sc Mehrdad, N., Mohapatra, H., Bagdouri, M., Chandran, P., Magnani, A., Cai,
  X., Puthenputhussery, A., Yadav, S., Lee, T., Zhai, C., et~al.}
\newblock Large language models for relevance judgment in product search.
\newblock {\em arXiv preprint arXiv:2406.00247\/} (2024).

\bibitem{moffat2007strategic}
{\sc Moffat, A., Webber, W., and Zobel, J.}
\newblock Strategic system comparisons via targeted relevance judgments.
\newblock In {\em Proceedings of the 30th annual international ACM SIGIR
  conference on research and development in information retrieval\/} (2007),
  pp.~375--382.

\bibitem{moghadasi2013low}
{\sc Moghadasi, S.~I., Ravana, S.~D., and Raman, S.~N.}
\newblock Low-cost evaluation techniques for information retrieval systems: A
  review.
\newblock {\em Journal of Informetrics 7}, 2 (2013), 301--312.

\bibitem{nogueira2020document}
{\sc Nogueira, R., Jiang, Z., and Lin, J.}
\newblock Document ranking with a pretrained sequence-to-sequence model.
\newblock {\em arXiv preprint arXiv:2003.06713\/} (2020).

\bibitem{nuray2006automatic}
{\sc Nuray, R., and Can, F.}
\newblock Automatic ranking of information retrieval systems using data fusion.
\newblock {\em Information processing \& management 42}, 3 (2006), 595--614.

\bibitem{oosterhuis2024reliable}
{\sc Oosterhuis, H., Jagerman, R., Qin, Z., Wang, X., and Bendersky, M.}
\newblock Reliable confidence intervals for information retrieval evaluation
  using generative ai.
\newblock In {\em Proceedings of the 30th ACM SIGKDD Conference on Knowledge
  Discovery and Data Mining\/} (2024), pp.~2307--2317.

\bibitem{pavlu2007practical}
{\sc Pavlu, V., and Aslam, J.}
\newblock A practical sampling strategy for efficient retrieval evaluation.
\newblock {\em College of Computer and Information Science, Northeastern
  University\/} (2007).

\bibitem{rahman2020efficient}
{\sc Rahman, M.~M., Kutlu, M., Elsayed, T., and Lease, M.}
\newblock Efficient test collection construction via active learning.
\newblock In {\em Proceedings of the 2020 ACM SIGIR on International Conference
  on Theory of Information Retrieval\/} (2020), pp.~177--184.

\bibitem{rahmani2024synthetic}
{\sc Rahmani, H.~A., Craswell, N., Yilmaz, E., Mitra, B., and Campos, D.}
\newblock Synthetic test collections for retrieval evaluation.
\newblock In {\em Proceedings of the 47th International ACM SIGIR Conference on
  Research and Development in Information Retrieval\/} (2024), pp.~2647--2651.

\bibitem{rahmani2024report}
{\sc Rahmani, H.~A., Siro, C., Aliannejadi, M., Craswell, N., Clarke, C.~L.,
  Faggioli, G., Mitra, B., Thomas, P., and Yilmaz, E.}
\newblock Report on the 1st workshop on large language model for evaluation in
  information retrieval (llm4eval 2024) at sigir 2024.
\newblock {\em arXiv preprint arXiv:2408.05388\/} (2024).

\bibitem{rahmani2024syndl}
{\sc Rahmani, H.~A., Wang, X., Yilmaz, E., Craswell, N., Mitra, B., and Thomas,
  P.}
\newblock Syndl: A large-scale synthetic test collection.
\newblock {\em arXiv preprint arXiv:2408.16312\/} (2024).

\bibitem{rajapakse2023improving}
{\sc Rajapakse, T.~C., and de~Rijke, M.}
\newblock Improving the generalizability of the dense passage retriever using
  generated datasets.
\newblock In {\em European Conference on Information Retrieval\/} (2023),
  Springer, pp.~94--109.

\bibitem{roitero2020effectiveness}
{\sc Roitero, K., Brunello, A., Serra, G., and Mizzaro, S.}
\newblock Effectiveness evaluation without human relevance judgments: A
  systematic analysis of existing methods and of their combinations.
\newblock {\em Information Processing \& Management 57}, 2 (2020), 102149.

\bibitem{roitero2018effectiveness}
{\sc Roitero, K., Soprano, M., and Mizzaro, S.}
\newblock Effectiveness evaluation with a subset of topics: A practical
  approach.
\newblock In {\em The 41st International ACM SIGIR Conference on Research \&
  Development in Information Retrieval\/} (2018), pp.~1145--1148.

\bibitem{sakai2007alternatives}
{\sc Sakai, T.}
\newblock Alternatives to bpref.
\newblock In {\em Proceedings of the 30th annual international ACM SIGIR
  conference on Research and development in information retrieval\/} (2007),
  pp.~71--78.

\bibitem{sakai2016topic}
{\sc Sakai, T.}
\newblock Topic set size design.
\newblock {\em Information Retrieval Journal 19}, 3 (2016), 256--283.

\bibitem{sander2021exam}
{\sc Sander, D.~P., and Dietz, L.}
\newblock Exam: How to evaluate retrieve-and-generate systems for users who do
  not (yet) know what they want.
\newblock In {\em DESIRES\/} (2021), pp.~136--146.

\bibitem{sanderson2010test}
{\sc Sanderson, M., et~al.}
\newblock Test collection based evaluation of information retrieval systems.
\newblock {\em Foundations and Trends{\textregistered} in Information Retrieval
  4}, 4 (2010), 247--375.

\bibitem{senter1967automated}
{\sc Senter, R., and Smith, E.~A.}
\newblock Automated readability index.
\newblock Tech. rep., Technical report, DTIC document, 1967.

\bibitem{thomas2024large}
{\sc Thomas, P., Spielman, S., Craswell, N., and Mitra, B.}
\newblock Large language models can accurately predict searcher preferences.
\newblock In {\em Proceedings of the 47th International ACM SIGIR Conference on
  Research and Development in Information Retrieval\/} (2024), pp.~1930--1940.

\bibitem{touvron2023llama}
{\sc Touvron, H., Lavril, T., Izacard, G., Martinet, X., Lachaux, M.-A.,
  Lacroix, T., Rozi{\`e}re, B., Goyal, N., Hambro, E., Azhar, F., et~al.}
\newblock Llama: Open and efficient foundation language models.
\newblock {\em arXiv preprint arXiv:2302.13971\/} (2023).

\bibitem{upadhyay2024llms}
{\sc Upadhyay, S., Kamalloo, E., and Lin, J.}
\newblock Llms can patch up missing relevance judgments in evaluation.
\newblock {\em arXiv preprint arXiv:2405.04727\/} (2024).

\bibitem{urbano2013measurement}
{\sc Urbano, J., Marrero, M., and Mart{\'\i}n, D.}
\newblock On the measurement of test collection reliability.
\newblock In {\em Proceedings of the 36th international ACM SIGIR conference on
  Research and development in information retrieval\/} (2013), pp.~393--402.

\bibitem{voorhees2000variations}
{\sc Voorhees, E.~M.}
\newblock Variations in relevance judgments and the measurement of retrieval
  effectiveness.
\newblock {\em Information processing \& management 36}, 5 (2000), 697--716.

\bibitem{robust2004}
{\sc Voorhees, E.~M.}
\newblock Overview of the trec 2004 robust retrieval track.

\bibitem{voorhees2009topic}
{\sc Voorhees, E.~M.}
\newblock Topic set size redux.
\newblock In {\em Proceedings of the 32nd international ACM SIGIR conference on
  Research and development in information retrieval\/} (2009), pp.~806--807.

\bibitem{voorhees2018building}
{\sc Voorhees, E.~M.}
\newblock On building fair and reusable test collections using bandit
  techniques.
\newblock In {\em Proceedings of the 27th ACM international conference on
  information and knowledge management\/} (2018), pp.~407--416.

\bibitem{voorhees2000overview}
{\sc Voorhees, E.~M., and Harman, D.}
\newblock Overview of the sixth text retrieval conference (trec-6).
\newblock {\em Information Processing \& Management 36}, 1 (2000), 3--35.

\bibitem{trec6}
{\sc Voorhees, E.~M., and Harman, D.}
\newblock Overview of the sixth text retrieval conference (trec-6).
\newblock {\em Information Processing \& Management 36}, 1 (2000), 3--35.

\bibitem{trec8}
{\sc Voorhees, E.~M., and Harman, D.~K.}
\newblock Overview of the eighth text retrieval conference (trec-8).
\newblock In {\em Text Retrieval Conference\/} (1999).

\bibitem{trec7}
{\sc Voorhees, E.~M., and Harman, D.~K.}
\newblock Overview of the seventh text retrieval conference (trec-7).

\bibitem{voorhees2022can}
{\sc Voorhees, E.~M., Soboroff, I., and Lin, J.}
\newblock Can old trec collections reliably evaluate modern neural retrieval
  models?
\newblock {\em arXiv preprint arXiv:2201.11086\/} (2022).

\bibitem{ye2022zerogen}
{\sc Ye, J., Gao, J., Li, Q., Xu, H., Feng, J., Wu, Z., Yu, T., and Kong, L.}
\newblock Zerogen: Efficient zero-shot learning via dataset generation.
\newblock In {\em Proceedings of the 2022 Conference on Empirical Methods in
  Natural Language Processing\/} (2022), pp.~11653--11669.

\bibitem{yu2024large}
{\sc Yu, Y., Zhuang, Y., Zhang, J., Meng, Y., Ratner, A.~J., Krishna, R., Shen,
  J., and Zhang, C.}
\newblock Large language model as attributed training data generator: A tale of
  diversity and bias.
\newblock {\em Advances in Neural Information Processing Systems 36\/} (2024).

\bibitem{yu2023regen}
{\sc Yu, Y., Zhuang, Y., Zhang, R., Meng, Y., Shen, J., and Zhang, C.}
\newblock Regen: Zero-shot text classification via training data generation
  with progressive dense retrieval.
\newblock In {\em Findings of the Association for Computational Linguistics:
  ACL 2023\/} (2023), pp.~11782--11805.

\bibitem{zhai2024large}
{\sc Zhai, C.}
\newblock Large language models and future of information retrieval:
  Opportunities and challenges.
\newblock In {\em Proceedings of the 47th International ACM SIGIR Conference on
  Research and Development in Information Retrieval\/} (2024), pp.~481--490.

\bibitem{zhang2024usimagent}
{\sc Zhang, E., Wang, X., Gong, P., Lin, Y., and Mao, J.}
\newblock Usimagent: Large language models for simulating search users.
\newblock In {\em Proceedings of the 47th International ACM SIGIR Conference on
  Research and Development in Information Retrieval\/} (2024), pp.~2687--2692.

\bibitem{zobel1998inverted}
{\sc Zobel, J., Moffat, A., and Ramamohanarao, K.}
\newblock Inverted files versus signature files for text indexing.
\newblock {\em ACM Transactions on Database Systems (TODS) 23}, 4 (1998),
  453--490.

\end{thebibliography}

\begin{appendices}

\setcounter{table}{0}
\renewcommand{\thetable}{A\arabic{table}}

\begin{table*}[!htb]
\scriptsize
\centering
  \begin{tabular}{ p{2cm} p{2.7cm} p{7.3cm} }
    \hline
    \textbf{Phase} & \textbf{Prompt} & \textbf{ChatGPT's response} \\ \hline
    Initialization Query  & $<$topic description$>$. Can you write a $<$document type$>$ about that? &  Environmental protection is a critical issue that affects the health and wellbeing of people and the planet as a whole. While many countries have taken significant steps to protect the environment and promote sustainable development, there are still countries that either do not practice or ignore environmental protective measures. ... \\ \hline
    
    Subtopic Generation Query  &  Can you write 100 subtopics related to this? Please be specific as much as possible. &  
    1) The impact of deforestation on the environment and local communities \newline
    2) The effects of oil and gas exploration on the environment \newline
    3) The impact of plastic pollution on marine life and ecosystems \newline
    4) The dangers of industrial pollution on air and water quality \newline
    %5) The effects of climate change on ecosystems and human populations \newline
    ... \\ \hline
    Document generation  &  Can you write a long text with a title about $<$subtopic$>$, within the scope of $<$topic description$>$ ? &  Title: The Devastating Impact of Deforestation on the Environment and Local Communities in Countries with Negligible Environmental Protective Measures \newline \newline
    Deforestation is the removal of trees from forests on a massive scale. It has been a critical issue in many countries, particularly those with weak environmental laws or enforcement mechanisms. Deforestation is a serious environmental problem that has far-reaching impacts on the planet and local communities. ... \\ \hline
    Random Non-Relevant Document Generation  &  Write me a $<$document type$>$ about any topic & Title: The Importance of Mental Health Awareness and Self-Care \newline \newline 
    Mental health is a critical aspect of our overall well-being. It refers to our psychological and emotional well-being and affects how we think, feel, and act. Mental health can be influenced by various factors, including genetics, environment, and lifestyle. It is essential to take care of our mental health, just like our physical health, to live a healthy and fulfilling life. ...\\ \hline
   Altered topic generation & Can you generate 10 variants of the next sentence by filling [MASK]: $<$ masked topic description $>$  
    \newline \newline
    Example: $<$topic description$>$
   & 
   1) Countries that do not practice or ignore public health protective measures. \newline
   2) Countries that do not practice or ignore workplace safety protective measures. \newline
   3) Countries that do not practice or ignore cybersecurity protective measures. \newline
   4) Countries that do not practice or ignore wildlife conservation protective measures. \newline
%   5) Countries that do not practice or ignore food safety protective measures. \newline
   ... \\ \hline
  \end{tabular}
  \caption{
ChatGPT prompts and corresponding responses for Topic 255 of TREC5. The topic description reads: "Countries that do not practice or ignore environmental protective measures". The masked version of the description is: "Countries that do not practice or ignore [MASK] protective measures.".} 
  \label{tab_llm_prompts}
\end{table*}

\end{appendices}

\end{document}